\documentclass[]{fairmeta}
\usepackage{amsthm,amsmath,amssymb}
\usepackage{graphicx}
\usepackage{natbib}
\usepackage[hidelinks,breaklinks,hypertexnames=False]{hyperref}
\usepackage{subfig}
\newtheorem{theorem}{Theorem}[]

\newtheorem{remark1}[theorem]{Remark}

\DeclareMathOperator{\E}{\mathop{}\mathbb{E}}

\renewcommand{\epsilon}{\varepsilon}

\title{Cumulative differences between subpopulations versus body mass index
in the Behavioral Risk Factor Surveillance System data}
\author[1]{Mark Tygert}
\affiliation[1]{Fundamental Artificial Intelligence Research at Meta}
\date{\today}
\correspondence{Mark Tygert at \email{mark@tygert.com}}

\begin{document}

\abstract{Prior works have demonstrated many advantages
of cumulative statistics over the classical methods of reliability diagrams,
ECEs (empirical, estimated, or expected calibration errors),
and ICIs (integrated calibration indices).
The advantages pertain to assessing calibration of predicted probabilities,
comparison of responses from a subpopulation to the responses
from the full population, and comparison of responses from one subpopulation
to those from a separate subpopulation.
The cumulative statistics include graphs of cumulative differences
as a function of the scalar covariate, as well as metrics
due to Kuiper and to Kolmogorov and Smirnov that summarize the graphs
into single scalar statistics and associated P-values
(also known as ``attained significance levels'' for significance tests).
However, the prior works have not yet treated data from biostatistics.

Fortunately, the advantages of the cumulative statistics extend
to the Behavioral Risk Factor Surveillance System (BRFSS)
of the Centers for Disease Control and Prevention.
This is unsurprising, since the mathematics is the same as in earlier works.
Nevertheless, detailed analysis of the BRFSS data is revealing
and corroborates the findings of earlier works.

Two methodological extensions beyond prior work that facilitate analysis
of the BRFSS are (1) empirical estimators of uncertainty for graphs
of the cumulative differences between two subpopulations,
such that the estimators are valid for any real-valued responses,
and (2) estimators of the weighted average treatment effect for the differences
in the responses between the subpopulations.
Both of these methods concern the case in which none
of the covariate's observed values for one subpopulation is equal
to any of the covariate's values for the other subpopulation.
The data analysis presented reports results for this case
as well as several others.
}

\maketitle

\section{Introduction}

A classic paradox popularized by~\cite{simpson} is that aggregating data
without consideration for a confounding covariate may result
in an empirical estimate of the average treatment effect (ATE)
having a sign opposite to the signs of the conditional ATEs
(which are simply the ATE conditioned on the observed values of the covariate).
This Simpson's Paradox can happen if the covariate is a confounder
and the distributions of the covariate are different
for the treated and untreated subpopulations, so that the averages
in the estimated ATE are taken over different distributions of the covariate.
In other words, the observed responses from one subpopulation
might look greater than the responses from another subpopulation
at the individual values of the covariate, while the average difference
in responses could be in the other direction when aggregating
without controlling for the confounding covariate, if the distributions
of the covariate for the different subpopulations are sufficiently different.
This poses a dilemma: aggregating across more values of the covariate
averages away randomness to make the estimates less noisy,
but then sacrifices the power to resolve effects at different values
of the covariate. There is a large literature studying such trade-offs,
particularly in the context of traditional methods for assessing calibration
and for assessing differences between subpopulations; the traditional methods
are the ICIs (integrated calibration indices) of~\cite{austin-steyerberg}
and the ECEs (empirical, estimated, or expected calibration errors)
and reliability diagrams reviewed by~\cite{brocker-smith}, \cite{brocker},
\cite{wilks}, and others.

Fortunately, cumulative statistics can avoid making any such explicit trade-off
between resolution and reliability when studying the differences in responses
from two subpopulations as a function of a scalar real-valued covariate.
The graph of the cumulative difference in responses between the subpopulations
displays the (weighted) average difference over any interval of the covariate
as the expected slope of the secant line connecting the two points
corresponding to the edges of the interval.
And an easily computed metric due to~\cite{kuiper} summarizes
into a single scalar statistic the differences across the full range
of the covariate; the Kuiper metric is the magnitude of the total difference
in responses between the subpopulations, totaled over the interval
of the covariate for which the magnitude of the total is greatest.
(The totals are normalized such that totaling over all possible values
of the covariate yields the weighted average treatment effect.)

A range of publications details various different kinds of comparisons.
Comparing a subpopulation to the full population
is treated by~\cite{tygert_full}.
Comparing two subpopulations directly, under the assumption
that the subpopulations' sets of observed values of the covariate are disjoint,
is treated by~\cite{tygert_two}.
Comparing two subpopulations directly, under the assumption
that each observed response from one subpopulation at a given value
of the covariate comes paired with exactly one response
from the other subpopulation at the same value of the covariate,
is treated by~\cite{kloumann-korevaar-mcconnell-tygert-zhao}.
Comparing a subpopulation to the special case in which the expected response
is a probability equal to the corresponding value of the covariate
is treated by~\cite{arrieta-ibarra-gujral-tannen-tygert-xu}
and~\cite{tygert_full}.
Efficient computational methods for calculating P-values
(also known as ``attained significance levels'') for formal significance tests
are detailed by~\cite{tygert_pvals}.
Section~\ref{methods} below reviews all this methodology
for the case of responses from subpopulations at values
of a single scalar covariate (in which case the values of the covariate
are known as ``scores'').
Extensions to conditioning on (also known as ``controlling for'')
multiple covariates are proposed by~\cite{tygert_multidim}.
A comprehensive treatment of assessing the calibration
of predicted probabilities that includes complete, rigorous proofs
of the statistical inferiority of the ECEs and ICIs was given
by~\cite{arrieta-ibarra-gujral-tannen-tygert-xu}; the present paper focuses
on comparing subpopulations, instead.

None of the prior publications mentioned in the previous paragraph
considers data from biostatistics.
That raises the question of whether the solutions provided in the prior works
apply to data from biomedical disciplines, too.
The present paper answers affirmatively: the mathematics is the same as before
and thus pertains just as well to a key biomedical data set,
namely, the Behavioral Risk Factor Surveillance System (BRFSS) of~\cite{brfss}.
Section~\ref{results} below reports empirical results with the BRFSS data.
Section~\ref{results} also gives indications that the methodology
should generalize to data from the social sciences, pointing
to data from~\cite{taylor-mickel} about expenditures
from California's Department of Developmental Services.

In order to handle the aforementioned data sets most completely, the present
paper makes two minor contributions beyond what prior work presented:
First, when two subpopulations' sets of observed values of a covariate
are disjoint from each other's, Subsection~\ref{disjoint} below provides
an empirical estimator in formula~(\ref{empirical}) of uncertainty
in the graph of cumulative differences between the subpopulations
(and the estimator is suitable for use in calculating P-values, too).
Second, Subsection~\ref{altATE} introduces simplified estimators
of the weighted average treatment effect
for the differences between the subpopulations' responses.
To be precise, the weighted average treatment effect (ATE)
is the weighted average of differences in responses
between the subpopulations, averaged over all values for the covariate.
Both contributions pertain to the case when the two subpopulations' sets
of values of a covariate are disjoint from each other's, which of course
happens precisely when none of the covariate's values
for one of the subpopulations is equal to any of the covariate's values
for the other subpopulation.

The rest of the present paper has the following structure: First,
Section~\ref{methods} sets notation and reviews the cumulative methodologies
of prior works. Section~\ref{methods} also introduces
the aforementioned two refinements, at the end of Subsection~\ref{disjoint}
and in Subsection~\ref{altATE}.
Then, Section~\ref{results} presents results of applying the methodology
to the BRFSS of~\cite{brfss} and points to software for analyzing the data
of~\cite{taylor-mickel}.\footnote{An open-source software package
that can automatically reproduce all results reported in the present paper
is available at \url{https://github.com/facebookresearch/cumbiostats}}
Finally, Section~\ref{conclusion} draws some conclusions.
Appendix~\ref{reliability_diagrams} reviews the construction
of the traditional reliability diagrams that Section~\ref{results} compares
to the cumulative methods.

\section{Methods}
\label{methods}

This section introduces the statistical methodology of the present paper.
Subsections~\ref{paired}, \ref{sub_vs_full}, and~\ref{disjoint}
review prior work on graphs and scalar summary statistics
of the cumulative differences (1) between two subpopulations whose observations
are paired, (2) between a subpopulation and the full population,
and (3) between two subpopulations whose sets of observed values
for their covariate are disjoint from each other's, respectively.
Subsection~\ref{disjoint} includes formula~(\ref{empirical}) ---
an estimator of uncertainty in the graphs and summary statistics
that can handle arbitrary real-valued responses,
unlike the estimators discussed in prior works.
Subsection~\ref{altATE} then presents an estimator
for the corresponding weighted average treatment effect that has
both advantages and disadvantages relative to prior propositions.
Appendix~\ref{reliability_diagrams} reviews reliability diagrams.

Before proceeding to the subsections, we set some notational conventions
used throughout.

For the covariate, we consider a set of distinct real-valued scalar scores,
$S_1 < S_2 < \dots < S_{\ell}$.
Each score $S_j^k = S_j$ comes paired with a scalar response $R_j^k$
and a positive scalar weight $W_j^k$,
for $j = 1$, $2$, \dots, $\ell$; and $k = 1$, $2$, \dots, $m_j$.
Thus, each score $S_j$ gets repeated $m_j$ times,
for $j = 1$, $2$, \dots, $\ell$.
The total number of observed triples $(S_j^k, R_j^k, W_j^k)$
is $m = m_1 + m_2 + \dots + m_{\ell}$,
as $j = 1$, $2$, \dots, $\ell$; and $k = 1$, $2$, \dots, $m_j$.

We consider subpopulations given by the pairs of indices
$i_1^0 < i_2^0 < \dots < i_{n_0}^0$ and
(in the case of two subpopulations)
$i_1^1 < i_2^1 < \dots < i_{n_1}^1$, where $i_j^0$ and $i_j^1$
are ordered pairs (of indices) ordered lexicographically.
The lexicographical ordering means that $i_j^0 = (p, q)$
is less than $i_k^0 = (r, s)$ if and only if
either $p < r$ or both $p = r$ and $q < s$.
By perturbing scores at random, we can ensure that each score becomes unique,
that is, $m_1 = m_2 = \dots = m_{\ell} = 1$
and so $m = m_1 + m_2 + \dots + m_{\ell} = \ell$.

When comparing a single subpopulation to the full population,
we set $n$ to be the number of different values that the scores take
for the subpopulation. There is no special need for every score to be distinct
when comparing a subpopulation to the full population
--- requiring $m_1 = m_2 = \dots = m_{\ell} = 1$ is unnecessary in this case.
So, when comparing a subpopulation to the full population, we need not perturb
the scores at random to ensure that all the scores be unique.
We omit such perturbations in the sequel when comparing a subpopulation
to the full population.

\begin{figure}
\begin{center}
\includegraphics[width=.53\textwidth]{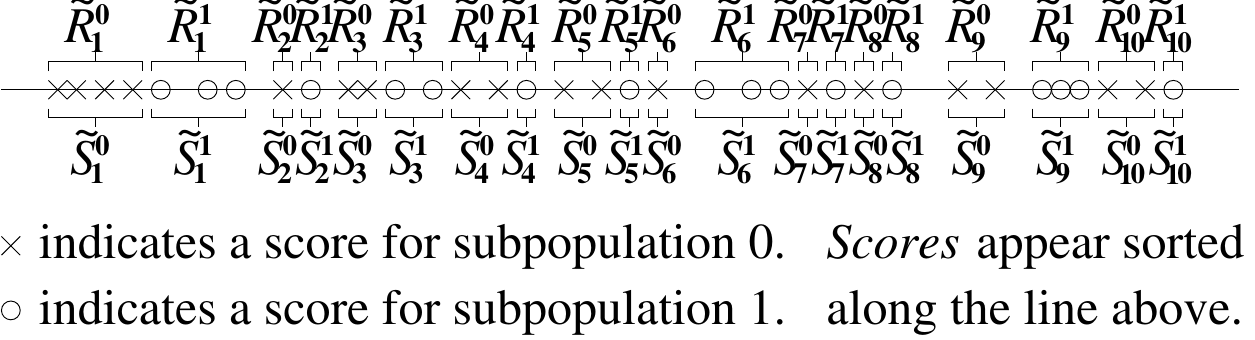}
\end{center}
\caption{When comparing directly two subpopulations in which no score
for one subpopulation is equal to any score for the other subpopulation,
we partition the original scores ($\times$'s and $\circ$'s)
as indicated and form the corresponding local weighted averages of scores
($\tilde{S}_1^0$, $\tilde{S}_2^0$, \dots, $\tilde{S}_n^0$;
$\tilde{S}_1^1$, $\tilde{S}_2^1$, \dots, $\tilde{S}_n^1$;
with $n = 10$ in this figure) and responses
($\tilde{R}_1^0$, $\tilde{R}_2^0$, \dots, $\tilde{R}_n^0$;
$\tilde{R}_1^1$, $\tilde{R}_2^1$, \dots, $\tilde{R}_n^1$;
with $n = 10$ in this figure). The corresponding local sums of weights
($\tilde{W}_1^0$, $\tilde{W}_2^0$, \dots, $\tilde{W}_n^0$;
$\tilde{W}_1^1$, $\tilde{W}_2^1$, \dots, $\tilde{W}_n^1$;
with $n = 10$ in this figure) also derive from the figure
according to the same groups as for the scores and responses.}
\label{partition}
\end{figure}

When comparing directly two subpopulations for which each score is distinct
from all others, we determine $n$ via an analogue of Figure~\ref{partition}
($n = 10$ in Figure~\ref{partition}).

In the case of paired samples, each score $S_j^k$ comes paired not only
with a scalar response $R_j^k$, but also with a scalar response $Q_j^k$,
for $j = 1$, $2$, \dots, $\ell$; and $k = 1$, $2$, \dots, $m_j$.
With paired samples, we take $n = \ell$.
The case of paired samples does not require all the scores to be unique
--- there is no need for $m_1 = m_2 = \dots = m_{\ell} = 1$
with paired samples.
There is no overwhelming advantage to perturbing the scores at random
to ensure their uniqueness for paired samples.
However, the case of paired samples does require that every score $S_j^k$
come with exactly two responses, $R_j^k$ and $Q_j^k$,
for $j = 1$, $2$, \dots, $\ell$; and $k = 1$, $2$, \dots, $m_j$.

Throughout, we assume that the responses are random variables
while the scores and weights are non-random
(which is equivalent to conditioning on the observed values of the scores
and weights). We assume that the responses are all jointly independent
as random variables.

\subsection{Paired samples}
\label{paired}

This subsection considers two subpopulations whose observations come together
in pairs, with each observed response from one subpopulation at a given score
paired with an observed response from the other subpopulation
at the same score. The present subsection simply reviews the work
of~\cite{kloumann-korevaar-mcconnell-tygert-zhao}
and~\cite{tygert_pvals}.

We form the weighted averages
\begin{equation}
\tilde{R}_j = \frac{\sum_{k=1}^{m_j} R_j^k \, W_j^k}{\sum_{k=1}^{m_j} W_j^k}
\end{equation}
and
\begin{equation}
\tilde{Q}_j = \frac{\sum_{k=1}^{m_j} Q_j^k \, W_j^k}{\sum_{k=1}^{m_j} W_j^k}
\end{equation}
for $j = 1$, $2$, \dots, $\ell$.
We also form the aggregated weights
\begin{equation}
\tilde{W}_j = \sum_{k=1}^{m_j} W_j^k
\end{equation}
for $j = 1$, $2$, \dots, $\ell$.

The accumulated weights are the aggregates $A_0 = 0$ and
\begin{equation}
A_j = \frac{\sum_{k=1}^j \tilde{W}_k}{\sum_{k=1}^{\ell} \tilde{W}_k}
\end{equation}
for $j = 1$, $2$, \dots, $\ell$.
The cumulative differences are $C_0 = 0$ and
\begin{equation}
C_j = \frac{\sum_{k=1}^j (\tilde{R}_k - \tilde{Q}_k) \, \tilde{W}_k}
           {\sum_{k=1}^{\ell} \tilde{W}_k}
\end{equation}
for $j = 1$, $2$, \dots, $\ell$.
We set $n = \ell$.

The expected slope of the secant line of the graph of $C_j$ versus $A_j$
from $j-1$ to $j$ is simply the expected difference
between the response $\tilde{R}_j$ from one subpopulation and the response
$\tilde{Q}_j$ from the other subpopulation:
\begin{equation}
\E\left[ \frac{C_j - C_{j-1}}{A_j - A_{j-1}} \right]
= \E[ \tilde{R}_j ] - \E[ \tilde{Q}_j ]
\end{equation}
for $j = 1$, $2$, \dots, $\ell$.
Thus, the slope of a secant line connecting two points on the graph
of $C_j$ versus $A_j$ becomes the weighted average difference
between the subpopulations over the long range of scores
between those two points where the secant line intersects the graph.

The weighted average treatment effect (ATE) is
\begin{equation}
\frac{\sum_{j=1}^{\ell} \sum_{k=1}^{m_j} (R_j^k - Q_j^k) \, W_j^k}
     {\sum_{j=1}^{\ell} \sum_{k=1}^{m_j} W_j^k} = C_{\ell},
\end{equation}
which is simply the ordinate (vertical coordinate) of the graph
of $C_j$ versus $A_j$ at the rightmost abscissa (horizontal coordinate).

Under the null hypothesis that the two subpopulations are drawn
independently from the same underlying probability distribution
whose mean is $\E[\tilde{R}_j] = \E[\tilde{Q}_j]$ at score $S_j$,
an estimate of the variance of $C_\ell$ is
\begin{equation}
\sigma^2 = \frac{\sum_{j=1}^{\ell}
                 (\tilde{R}_j - \tilde{Q}_j)^2 \, (\tilde{W}_j)^2}
                {\Bigl( \sum_{j=1}^{\ell} \tilde{W}_j \Bigr)^2}.
\end{equation}
Displaying at the origin of the plot a triangle whose tip-to-tip height
is $4\sigma$ indicates the expected deviation over the full range of scores
at roughly the 95\% confidence level.

There are two natural statistics that summarize into single scalars the plots
of cumulative differences. Underlying both metrics is the fact that,
under the null hypothesis of no deviation between the subpopulations, 
the cumulative differences would follow a driftless random walk.
Thus, under the null hypothesis, the graph of cumulative differences
would stay reasonably flat, not deviating significantly away from 0.

One metric is due to~\cite{kuiper}, the range of deviations:
\begin{equation}
{\rm Kuiper} = \max_{0 \le j \le \ell} C_j - \min_{0 \le j \le \ell} C_j,
\end{equation}
where $C_0 = 0$.
An alternative expression for the Kuiper statistic is
\begin{equation}
\label{intuitive_pair}
{\rm Kuiper} = \max_{1 \le j \le k \le \ell}
\left| \frac{\sum_{i=j}^k (\tilde{R}_i - \tilde{Q}_i) \, \tilde{W}_i}
            {\sum_{i=1}^{\ell} \tilde{W}_i} \right|,
\end{equation}
which is simply the absolute value of the total weighted deviation
between the subpopulations, totaled over the interval of scores
for which the magnitude of the total is greatest.

Another natural metric is due to~\cite{kolmogorov} and~\cite{smirnov},
the maximum of the absolute values of deviations:
\begin{equation}
\hbox{Kolmogorov-Smirnov} = \max_{1 \le j \le \ell} |C_j|,
\end{equation}
which unfortunately lacks the intuitive interpretation
of~(\ref{intuitive_pair}).

\subsection{Deviation of a subpopulation from the full population}
\label{sub_vs_full}

This subsection compares a subpopulation to the full population,
reviewing the work of~\cite{tygert_full} and~\cite{tygert_pvals}.

We form the weighted averages
\begin{equation}
\tilde{R}_j = \frac{\sum_{k\,:\,(i_k^0)_1 = j}
                    R_j^{(i_k^0)_2} \, W_j^{(i_k^0)_2}}
                   {\sum_{k\,:\,(i_k^0)_1 = j} W_j^{(i_k^0)_2}}
\end{equation}
and
\begin{equation}
\tilde{Q}_j = \frac{\sum_{i\,:\,B_{j-1} < S_i \le B_j} \sum_{k=1}^{m_i}
                    R_j^k \, W_j^k}
                   {\sum_{i\,:\,B_{j-1} < S_i \le B_j} \sum_{k=1}^{m_i} W_j^k}
\end{equation}
for $j = 1$, $2$, \dots, $\ell$,
where
\begin{equation}
B_j = \frac{S_j + S_{j+1}}{2}
\end{equation}
for $j = 1$, $2$, \dots, $\ell - 1$,
\begin{equation}
B_0 = -\infty,
\end{equation}
and
\begin{equation}
B_{\ell} = \infty.
\end{equation}
We also form the aggregated weights
\begin{equation}
\tilde{W}_j = \sum_{k\,:\,(i_k^0)_1 = j} W_j^{(i_k^0)_2}
\end{equation}
for $j = 1$, $2$, \dots, $\ell$.

As in Subsection~\ref{paired}, the accumulated weights are the aggregates
$A_0 = 0$ and
\begin{equation}
A_j = \frac{\sum_{k=1}^j \tilde{W}_k}{\sum_{k=1}^{\ell} \tilde{W}_k}
\end{equation}
for $j = 1$, $2$, \dots, $\ell$,
and the cumulative differences are $C_0 = 0$ and
\begin{equation}
C_j = \frac{\sum_{k=1}^j (\tilde{R}_k - \tilde{Q}_k) \, \tilde{W}_k}
           {\sum_{k=1}^{\ell} \tilde{W}_k}
\end{equation}
for $j = 1$, $2$, \dots, $\ell$.
We set $n$ to be the number of different values that the scores take
for the subpopulation.

The expected slope of the secant line of the graph of $C_j$ versus $A_j$
from $j-1$ to $j$ is simply the expected difference
between the response $\tilde{R}_j$ from the subpopulation and the response
$\tilde{Q}_j$ from the full population:
\begin{equation}
\E\left[ \frac{C_j - C_{j-1}}{A_j - A_{j-1}} \right]
= \E[ \tilde{R}_j ] - \E[ \tilde{Q}_j ]
\end{equation}
for $j = 1$, $2$, \dots, $\ell$.
Thus, the slope of a secant line connecting two points on the graph
of $C_j$ versus $A_j$ becomes the weighted average difference
between the subpopulation and the full population over the long range
of scores between those two points where the secant line intersects the graph.

The weighted average treatment effect (ATE) is
\begin{equation}
\frac{\sum_{j=1}^{\ell} (\tilde{R}_j - \tilde{Q}_j) \, \tilde{W}_j}
     {\sum_{j=1}^{\ell} \tilde{W}_j} = C_{\ell},
\end{equation}
which is just the ordinate (vertical coordinate) of the graph
of $C_j$ versus $A_j$ at the rightmost abscissa (horizontal coordinate).

Under the null hypothesis that the subpopulation and full population are drawn
independently from the same underlying Bernoulli distribution,
an estimate of the variance of $C_{\ell}$ is
\begin{equation}
\sigma^2 = \frac{\sum_{j=1}^{\ell} \tilde{Q}_j (1 - \tilde{Q}_j)
                 \sum_{k\,:\,(i_k^0)_1 = j} \Bigl( W_j^{(i_k^0)_2} \Bigr)^2}
{\Bigl( \sum_{j=1}^{\ell} \sum_{k\,:\,(i_k^0)_1 = j} W_j^{(i_k^0)_2} \Bigr)^2},
\end{equation}
assuming that there are many observations from the full population
close in score to that for each observation from the subpopulation
(so $\tilde{Q}_j \approx \E[\tilde{Q}_j]$).
Displaying at the origin of the plot a triangle whose tip-to-tip height
is $4\sigma$ indicates the expected deviation over the full range of scores
at about the 95\% confidence level.

As in the other subsections, there are two standard statistics
that summarize into single scalars the graphs of cumulative differences.
As before, under the null hypothesis of no deviation
between the subpopulation and the full population, 
the cumulative differences would follow a driftless random walk.
So, under the null hypothesis, the graph of cumulative differences would be
reasonably flat and horizontal and not deviate significantly away from 0.

One scalar summary statistic is due to~\cite{kuiper}, the range of deviations:
\begin{equation}
{\rm Kuiper} = \max_{0 \le j \le \ell} C_j - \min_{0 \le j \le \ell} C_j,
\end{equation}
where $C_0 = 0$.
An alternative expression for the Kuiper statistic is
\begin{equation}
\label{intuitive_subpop}
{\rm Kuiper} = \max_{1 \le j \le k \le \ell}
\left| \frac{\sum_{i=j}^k (\tilde{R}_i - \tilde{Q}_i) \, \tilde{W}_i}
            {\sum_{i=1}^{\ell} \tilde{W}_i} \right|,
\end{equation}
which is simply the absolute value of the total weighted deviation
between the subpopulation and the full population,
totaled over the interval of scores for which the absolute value
of the total is greatest.

The other standard metric is due to~\cite{kolmogorov} and~\cite{smirnov},
the maximum of the absolute values of deviations:
\begin{equation}
\hbox{Kolmogorov-Smirnov} = \max_{1 \le j \le \ell} |C_j|,
\end{equation}
which unfortunately lacks the intuitive interpretation
of~(\ref{intuitive_subpop}).

\subsection{Deviation between two subpopulations
            whose sets of scores are disjoint}
\label{disjoint}

This subsection considers two subpopulations whose sets of scores are disjoint
from each other's, reviewing the work of~\cite{tygert_two}.
The present subsection also introduces an empirical estimator of uncertainty
in the graphs of cumulative differences of responses
between the two subpopulations and in the associated scalar summary statistics;
this estimator of~(\ref{empirical}) below is valid
for any real-valued responses, not limited to responses
from Bernoulli distributions as in earlier works.
The present subsection combines this estimator with the approach
of~\cite{tygert_pvals}, too.

Recall that the scores are sorted such that $S_1 < S_2 < \dots < S_{\ell}$.
For comparing two general (not necessarily paired
as in Subsection~\ref{paired}) subpopulations directly,
we require that none of the scores for one of the subpopulations
be equal to any of the scores for the other subpopulation; we can ensure this
(that the sets of scores for the subpopulations are disjoint
from each other's) by randomly perturbing the scores if necessary.
We can then sort all the indices for both subpopulations combined,
as in Figure~\ref{partition}
(reversing the labels 0 and 1 for the subpopulations if necessary).
As shown in the figure, we group together the largest contiguous ranges
of scores for the subpopulations that do not intersect the other ranges.
We denote by $\tilde{S}_1^0$ the weighted average
of all scores in the leftmost group for subpopulation 0;
we denote by $\tilde{S}_1^1$ the weighted average
of all scores in the leftmost group for subpopulation 1.
We denote by $\tilde{S}_2^0$ the weighted average
of all scores in the second to leftmost group for subpopulation 0;
we denote by $\tilde{S}_2^1$ the weighted average
of all scores in the second to leftmost group for subpopulation 1.
More generally, we denote by $\tilde{S}_j^0$ the weighted average
of all scores in the $j$th to leftmost group for subpopulation 0;
we denote by $\tilde{S}_j^1$ the weighted average
of all scores in the $j$th to leftmost group for subpopulation 1.
By construction, we obtain
$\tilde{S}_1^0 < \tilde{S}_1^1 < \tilde{S}_2^0 < \tilde{S}_2^1 < \dots
< \tilde{S}_n^0 < \tilde{S}_n^1$,
where $n$ is the total number of groups for each subpopulation
(for simplicity of exposition, we assume that there are $n$ groups for each;
another possibility is that subpopulation~1 has only $n-1$ groups).

For every group of scores, we denote the associated weighted averages
of responses by $\tilde{R}_j^k$, so that $\tilde{R}_j^k$ is
the weighted average of responses corresponding
to the weighted average score $\tilde{S}_j^k$,
for $j = 1$, $2$, \dots, $n$; and $k = 0$, $1$.
Similarly, for every group of scores, we denote the associated sum
of weights by $\tilde{W}_j^k$, so that $\tilde{W}_j^k$ is
the sum of the weights corresponding
to the weighted average score $\tilde{S}_j^k$
for $j = 1$, $2$, \dots, $n$; and $k = 0$, $1$.

\begin{figure}
\begin{center}
\subfloat[odd diffs.\ ($D_{2k-1}$)\label{odddiffs}]{
\makebox[1.6\width]{\includegraphics[height=.15\textheight]{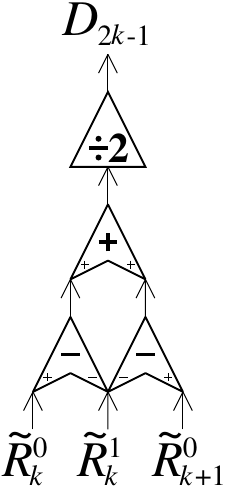}}
}
\quad\quad\quad\quad
\subfloat[even diffs.\ ($D_{2k}$)\label{evendiffs}]{
\makebox[1.5\width]{\includegraphics[height=.15\textheight]{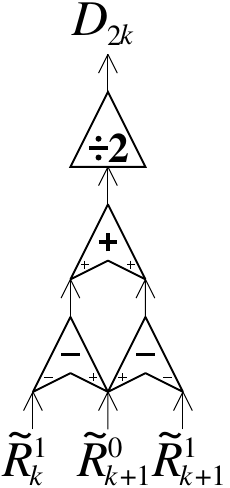}}
}
\quad\quad\quad\quad
\subfloat[odd sums ($T_{2k-1}$)\label{oddsums}]{
\makebox[1.5\width]{\includegraphics[height=.15\textheight]{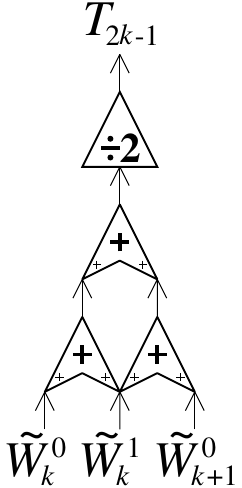}}
}
\quad\quad\quad\quad
\subfloat[even sums ($T_{2k}$)\label{evensums}]{
\makebox[1.5\width]{\includegraphics[height=.15\textheight]{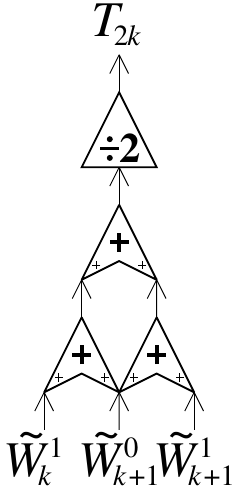}}
}
\end{center}
\caption{When comparing directly two subpopulations whose sets of scores
are disjoint from each other's, we form differences and sums as indicated,
using the local weighted averages of responses
($\tilde{R}_1^0$, $\tilde{R}_2^0$, \dots, $\tilde{R}_n^0$;
$\tilde{R}_1^1$, $\tilde{R}_2^1$, \dots, $\tilde{R}_n^1$)
and associated local sums of weights
($\tilde{W}_1^0$, $\tilde{W}_2^0$, \dots, $\tilde{W}_n^0$;
$\tilde{W}_1^1$, $\tilde{W}_2^1$, \dots, $\tilde{W}_n^1$)
from Figure~\ref{partition}.
}
\label{diffsums}
\end{figure}

We then form the averages of the weighted average forward and backward
differences according to Figures~\ref{odddiffs} and~\ref{evendiffs}.
(Of course, the average of the forward and backward differences
is a central, centered difference.)
We also form the corresponding totals of the weights according
to Figures~\ref{oddsums} and~\ref{evensums}.
Finally, we form the weighted cumulative differences $C_0 = 0$ and
\begin{equation}
C_j = \frac{\sum_{k=1}^j D_k \, T_k}{\sum_{k=1}^{2n-2} T_k}
\end{equation}
for $j = 1$, $2$, \dots, $2n-2$.
We also form the accumulated aggregate weights $A_0 = 0$ and
\begin{equation}
A_j = \frac{\sum_{k=1}^j T_k}{\sum_{k=1}^{2n-2} T_k}
\end{equation}
for $j = 1$, $2$, \dots, $2n-2$.

The expected slope of the secant line of the graph of $C_j$ versus $A_j$
from $j-1$ to $j$ is simply the expected centered difference $D_j$
between the responses of the subpopulations being compared:
\begin{equation}
\E\left[ \frac{C_j - C_{j-1}}{A_j - A_{j-1}} \right] = \E[ D_j ]
\end{equation}
for $j = 1$, $2$, \dots, $2n-2$.
Thus, the slope of a secant line connecting two points on the graph
of $C_j$ versus $A_j$ becomes the weighted average difference
between the subpopulations over the long range of scores
between those two points where the secant line intersects the graph.

Under the null hypothesis that the subpopulations are drawn
independently from the same underlying probability distribution,
an estimate of the variance of $C_{2n-2}$ is
\begin{equation}
\label{empirical}
\sigma^2 = \frac{\sum_{j=1}^{2n-1} (D_{j-1} - D_j)^2 \, (T_{j-1} + T_j)^2}
                {4 \Bigl( \sum_{j=1}^{2n-2} T_j \Bigr)^2},
\end{equation}
which includes a factor of 2 to adjust for the dependence between
the even and odd entries of the sequence $D_1$, $D_2$, \dots, $D_{2n-2}$
and with $D_0 = D_{2n-1} = T_0 = T_{2n-1} = 0$;
the factor of 4 in the denominator compensates for including the weights
twice prior to squaring them, and the aforementioned factor of 2
cancels with the need to divide by 2 to compensate
for doubling the variance in the difference ($D_{j-1} - D_j$)
if $D_{j-1}$ and $D_j$ were independent.
Notice how any linear trend will cancel in the difference ($D_{j-1} - D_j$)
from~(\ref{empirical}), with the minor drawback of decreasing the resolution
of the estimator, spreading the estimate across successive entries
$j-1$ and $j$ in the sequence.
Displaying at the origin of the plot a triangle whose tip-to-tip height
is $4\sigma$ indicates the expected deviation over the full range of scores
at roughly the 95\% confidence level.

As in the previous subsections, there are two natural metrics
that summarize into single scalars the graphs of cumulative differences.
As before, under the null hypothesis of no deviation
between the two subpopulations,
the cumulative differences would form a driftless random walk.
Thus, under the null hypothesis, the graph of cumulative differences
would stay flat and horizontal and not deviate significantly away from 0.

One summary statistic is due to~\cite{kuiper}, the range of deviations:
\begin{equation}
{\rm Kuiper} = \max_{0 \le j \le 2n-2} C_j - \min_{0 \le j \le 2n-2} C_j,
\end{equation}
where $C_0 = 0$.
An alternative expression for the Kuiper statistic is
\begin{equation}
\label{intuitive_subpops}
{\rm Kuiper} = \max_{1 \le j \le k \le 2n-2}
\left| \frac{\sum_{i=j}^k D_i \, T_i}{\sum_{i=1}^{2n-2} T_i} \right|,
\end{equation}
which is simply the absolute value of the total weighted deviation
between the subpopulations,
totaled over the interval of scores for which the absolute value
of the total is greatest.

The other summary statistic is due to~\cite{kolmogorov} and~\cite{smirnov},
the maximum of the absolute values of deviations:
\begin{equation}
\hbox{Kolmogorov-Smirnov} = \max_{1 \le j \le 2n-2} |C_j|,
\end{equation}
which unfortunately lacks the intuitive interpretation
of~(\ref{intuitive_subpops}).

There are two possible definitions of the weighted average treatment effect
(ATE) for this case of comparing directly two subpopulations whose sets
of scores are disjoint. The first is simply ATE = $C_{2n-2}$,
as discussed in earlier works. The results in the remainder
of the present paper report a second possibility,
detailed in the following subsection.

\subsection{Weighted average treatment effect between two subpopulations
whose sets of scores are disjoint}
\label{altATE}

This subsection introduces an alternative to the estimator
of the weighted average treatment effect from the previous subsection,
Subsection~\ref{disjoint}. The alternative has both advantages
and disadvantages.

For every score $S$, we calculate the difference of the corresponding response
from the response of the other subpopulation at the greatest score less than
$S$ for which the other subpopulation has a response. We also calculate
the difference of the same corresponding response from the response
of the other subpopulation at the least score greater than $S$
for which the other subpopulation has a response.
We average these two differences for each score $S$
(making sure to subtract the response for subpopulation 1
from the response of subpopulation 0 in all differences).
We compute the weighted average of all such average differences,
weighted by the normalized weight associated with the score $S$
(normalizing the weights such that their sum is 1/2 for each subpopulation).
This weighted average yields our weighted average treatment effect (ATE).

This ATE differs from the alternative ($C_{2n-2}$) mentioned
in the previous subsection mainly in the computation of differences between
adjacent contiguous groups of scores, where each group corresponds to responses
from only one of the subpopulations, as Figure~\ref{partition} illustrates.
In the previous subsection, the difference is between the weighted averages
over all scores in a group. In the present subsection,
the difference is between the weighted average over all scores in one group
from only the absolute nearest responses of the other subpopulation
(rather than from the weighted averages over all responses
of the other subpopulation in the neighboring groups).
Here, ``nearest'' refers to proximity of the corresponding scores.

If the uniqueness of each score gets ensured via randomly perturbing
the original scores, then the ATE will vary a bit at random.
The ATE proposed in the present subsection is less stable
to the random perturbations than the ATE defined in the previous subsection.
The ATE proposed here is simpler to define, however.

\section{Results and discussion}
\label{results}

This section applies the methods of the previous section,
Section~\ref{methods}, to the analysis of data
from the Behavioral Risk Factor Surveillance System (BRFSS),
as described by~\cite{brfss}.\footnote{Permissively licensed
open-source software which can automatically reproduce all results
reported here is available at
\url{https://github.com/facebookresearch/cumbiostats}}
Subsection~\ref{cdcbrfss} treats the BRFSS data;
Subsubsection~\ref{brfss_discussion} discusses results from the BRFSS
in detail.
Subsection~\ref{dds_taylor-mickel} provides pointers to an automated script
for processing the popular data from~\cite{taylor-mickel}
about California's Department of Developmental Services.
Reliability diagrams (reviewed in Appendix~\ref{reliability_diagrams})
are standard, yet look to be inferior in all the results.

\subsection{Centers for Disease Control's
Behavioral Risk Factor Surveillance System (BRFSS)}
\label{cdcbrfss}

This subsection applies the methods of the previous section
to a weighted data set from the BRFSS
of~\cite{brfss}.\footnote{Public-domain data and documentation for the BRFSS
is available from the Centers for Disease Control and Prevention at
\url{https://www.cdc.gov/brfss/data_documentation/index.htm}}
Health departments of states in the United States regularly conduct surveys
via telephone (both landline and cellular), coordinated by the BRFSS
of the Centers for Disease Control and Prevention,
which the latter aggregate and anonymize.
Inter alia, the data include responses to yes-or-no questions
about health problems that the participants in the survey may have or have had,
such as those indicated in the captions
to Figures~\ref{angina_heart-attack}--\ref{hiv_kidney3}.
The data pair the responses to those yes-or-no questions
with the body mass index (BMI) of each individual participant,
together with a sampling weight. The data also report the individuals' heights.

All analysis presented here encodes having or having had a given health problem
with the number 1 and encodes never having had the health problem
with the number 0. The BMI is the ratio of the individual's weight in kilograms
to the square of the individual's height in meters.
The data comes from 2022 and includes 396,326 participants with BMIs reported
(this filters down from all 445,132 to only those with BMIs).

In Figures~\ref{angina_heart-attack}--\ref{hiv_kidney3}
for the present subsection, the scores are the BMIs
(randomly perturbed when indicated).
In Figures~\ref{men_women1}--\ref{men_women3}, the scores are the heights
in centimeters, randomly perturbed such that each individual's height
becomes distinct from all others'.

Figures~\ref{angina_heart-attack} and~\ref{stroke_kidney-disease}
compare paired samples, as in Subsection~\ref{paired}.
In Figure~\ref{angina_heart-attack}, each observed pair of responses
consists of whether the corresponding individual has angina
or coronary heart disease and whether the individual has had a heart attack.
In Figure~\ref{stroke_kidney-disease}, each observed pair of responses
consists of whether the corresponding individual has had a stroke
and whether the individual has kidney disease.

Figure~\ref{heart-attack_doctor} compares a single subpopulation
to the full population, as in Subsection~\ref{sub_vs_full}.
The responses in Figure~\ref{heart-attack_doctor}
are whether the corresponding individuals could afford to see a doctor.
The subpopulation consists of those who have had a heart attack.

Figures~\ref{hiv_kidney1}--\ref{men_women3} compare two subpopulations
as in Subsection~\ref{disjoint}, perturbing their scores at random
to ensure that every score is distinct.

In Figures~\ref{hiv_kidney1}--\ref{hiv_kidney3},
one subpopulation consists of those who have been tested for HIV,
while the other consists of those who have not been tested;
the responses are whether the corresponding individuals have kidney disease.
As with the earlier figures, the scores
for Figures~\ref{hiv_kidney1}--\ref{hiv_kidney3} are the BMIs.
Each of Figures~\ref{hiv_kidney1}--\ref{hiv_kidney3}
perturbs the scores using a different seed for the random number generator.

In Figures~\ref{men_women1}--\ref{men_women3},
one subpopulation is men and the other is women.
The responses are the BMIs; the corresponding values of the covariates
(used as scores) are individuals' heights in centimeters.
This analyzes the dependency of BMI on height.
These figures provide examples in which the responses can take on real values
other than just 0 and 1.
Each of Figures~\ref{men_women1}--\ref{men_women3}
uses a different random seed for perturbing the scores.

\subsubsection{Discussion}
\label{brfss_discussion}

The cumulative graph in Figure~\ref{angina_heart-attack} reveals that
having had a heart attack is significantly and fairly uniformly more likely
than having angina or coronary heart disease for those with BMI 21 or less.
Discerning that range (less than BMI of 21) of roughly uniform discrepancy
in the prevalence is difficult from the traditional reliability diagrams
displayed in Figure~\ref{angina_heart-attack}.
According to Figure~\ref{angina_heart-attack},
having angina or coronary heart disease is significantly more likely
than having had a heart attack for those with BMI greater than 40.
Thus, having the chronic disease (angina or coronary heart disease)
is more likely than having had the acute disease (a heart attack)
for the morbidly obese, while the opposite is the case for the very thin.

The cumulative graph in Figure~\ref{stroke_kidney-disease} makes clear that
having kidney disease is significantly and fairly uniformly more likely
than having had a stroke for those with BMI 35 or greater.
Discerning that the discrepancy in the prevalence is fairly uniform
over that range (BMI greater than 35) is really hard
using only the traditional reliability diagrams shown
in Figure~\ref{stroke_kidney-disease}.
That kidney disease is so prevalent among the highly obese is unsurprising,
yet knowing that the increased likelihood is roughly uniform
for BMI greater than 35 may be of some interest.

The reliability diagrams in Figure~\ref{heart-attack_doctor}
are difficult to reconcile with each other for BMI around 30
(Simpson's Paradox looks to be part of the reason).
The cumulative graph makes clear that the underlying cause
is a slight aberration from the general trend of dependence on increasing BMI
for BMI near 30. Figure~\ref{heart-attack_doctor} shows that those who have had
a heart attack are more likely to have been unable to afford to see a doctor
than the full population (the full population consists of both those who have
and those who have not had a heart attack) for all BMIs, but especially
for BMI less than 20. However, accurately quantifying the increased correlation
for BMI less than 20 is hard using only the reliability diagrams.
The cumulative graphs facilitate quantification for the very thin
of how much more correlated those who have had a heart attack are
with being unable to afford to see a doctor, which could be of interest
in epidemiology. (Needless to say, estimating any such correlations comes
with the caveat that survey participants may not know or tell the truth).

Figures~\ref{hiv_kidney1}--\ref{hiv_kidney3} illustrate that those tested
for HIV are more correlated with having been diagnosed with kidney disease
than those not tested, at least among those with BMI greater than 25.
However, the cumulative graphs make clear that those tested for HIV are
actually much less correlated with having been diagnosed with kidney disease
than those not tested for HIV, among those with BMI less than 20.
The reliability diagrams show the same phenomenon, but quantifying
the size of the difference in correlations is tricky using only
the reliability diagrams --- the results of the quantification depend strongly
on the numbers of bins in the reliability diagrams.

The cumulative graphs in Figures~\ref{men_women1}--\ref{men_women3}
reveal that men have higher BMI than women of the same height for all heights,
but especially so for the few very, very short people.
(If the reported heights are accurate, then such very short people probably
are largely dwarfs and other little people with developmental disabilities;
note that all individuals in the survey must be 18 years or older
in order to participate --- all participants are adults.)
Quantifying just how much greater BMI is for men than for women
among the very, very short is very easy via the slope of the cumulative graphs,
whereas quantification using the corresponding reliability diagrams
is quixotic at best.

\subsection{California's Department of Developmental Services}
\label{dds_taylor-mickel}

A particularly popular data set was generated by~\cite{taylor-mickel}
to reflect raw data from the Department of Developmental Services
of California.\footnote{Data from~\cite{taylor-mickel} is available
at \url{http://www.StatLit.org/XLS/2014-Taylor-Mickel-Paradox-Data.xlsx}
and documentation is available
at \url{http://jse.amstat.org/v22n1/mickel/paradox_documentation.docx}}
Footnote~1 of~\cite{taylor-mickel} states,
``The data set originated from DDS's Client Master File.
In order to remain in compliance with California State Legislation,
the data have been altered to protect the rights and privacy
of specific individual consumers.''
An automated Python script for downloading and processing the data
that \cite{taylor-mickel} generated is available in the associated repository
from GitHub.\footnote{A script that automatically downloads and processes
the data set of~\cite{taylor-mickel} is available in the Python file,
{\tt tm.py}, in a subdirectory, {\tt codes}, of the repository available at
\url{https://github.com/facebookresearch/cumbiostats}}
The data includes the age, ethnicity/race (Asian, Black, Hispanic, or White),
and the total expenditures in dollars that the state made in a year
for each individual in the data set that \cite{taylor-mickel} synthesized.

\newlength{\imsize}
\setlength{\imsize}{.38\textwidth}
\newlength{\fudger}
\setlength{\fudger}{1.6in}

\begin{figure}
\begin{center}
\subfloat{
\includegraphics[width=\imsize]
{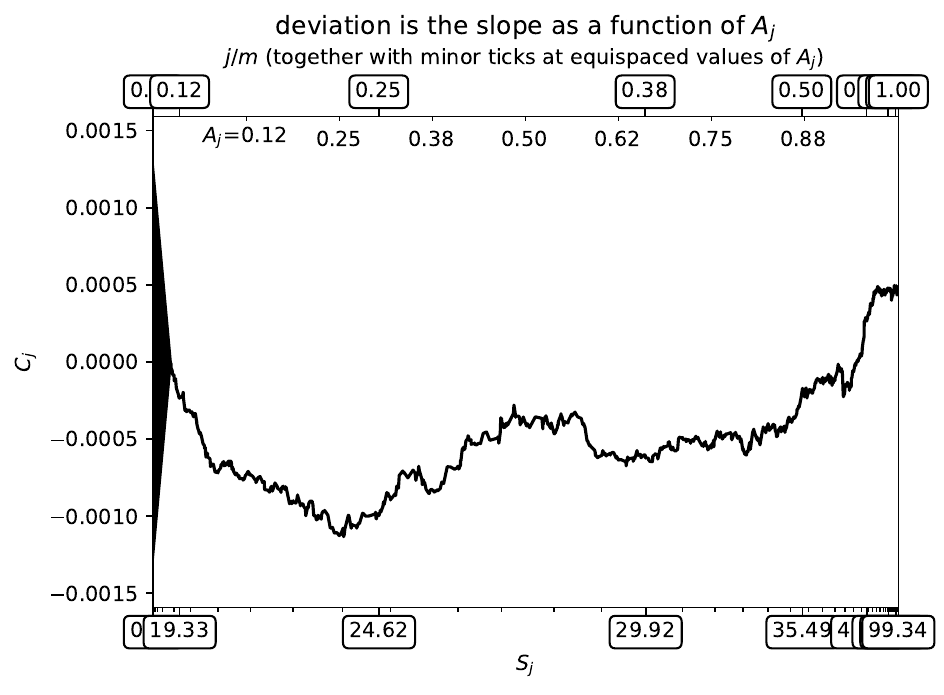}
\parbox{\imsize}{\footnotesize
$m =$ 396,326 (with $\ell =$ 3,985 distinct scores) \\
$n =$ 3,985 \\
Kuiper's statistic $= 0.001629 / \sigma = 2.456$; the asymptotic P-value
$= 0.05622$ \\
Kolmogorov-Smirnov's $= 0.001132 / \sigma = 1.707$; asymptotic P-value
$= 0.1755$ \\
ATE $= 0.0004812 / \sigma = 0.7254$
\vspace{\fudger}
}
}

\vspace{-.7\fudger}

\subfloat{
\includegraphics[width=\imsize]
{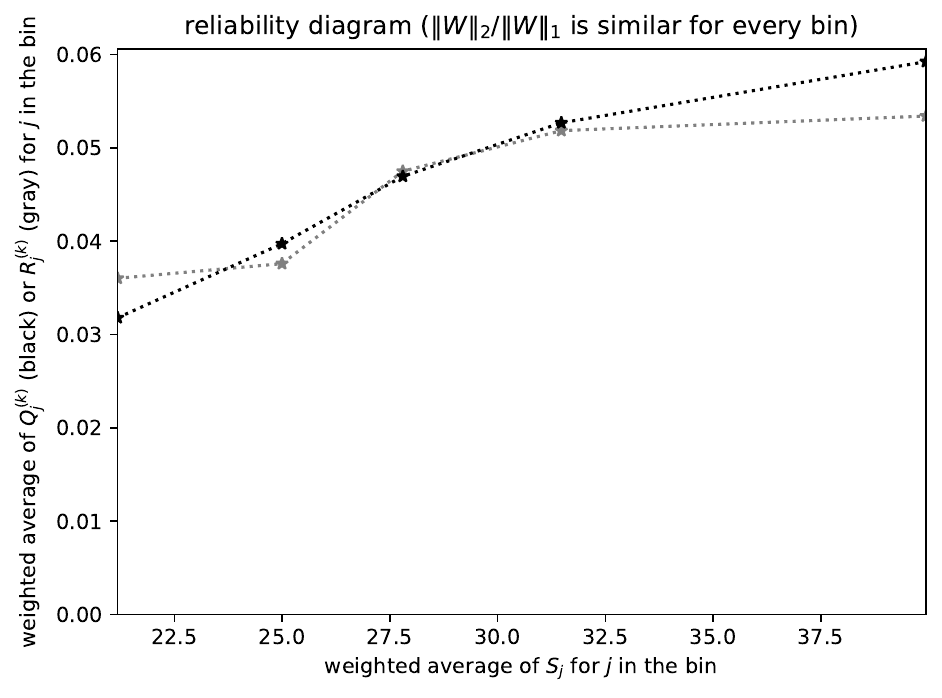}
}
\subfloat{
\includegraphics[width=\imsize]
{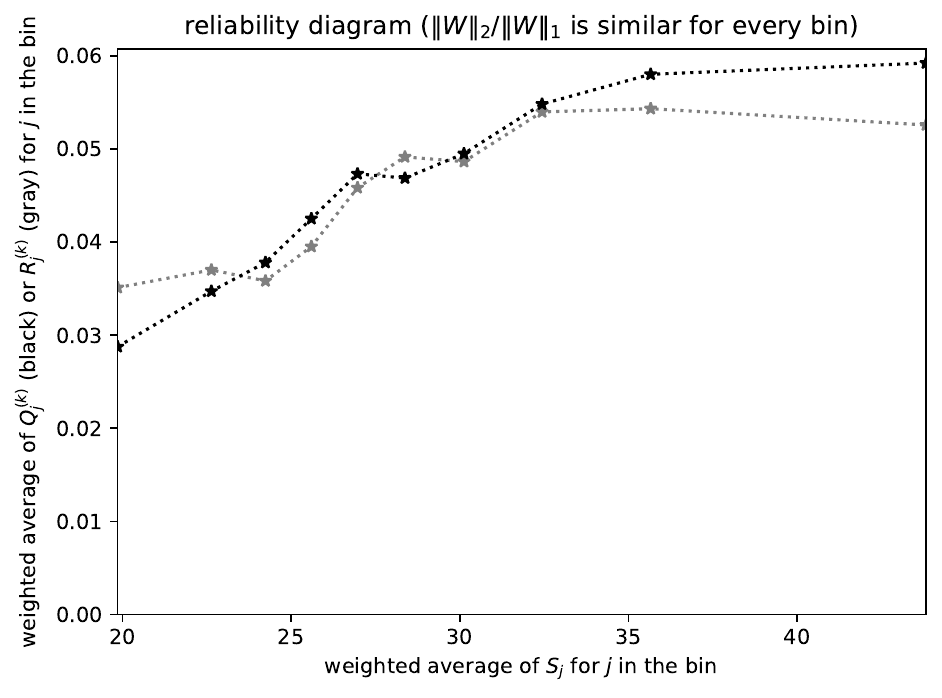}
}

\subfloat{
\includegraphics[width=\imsize]
{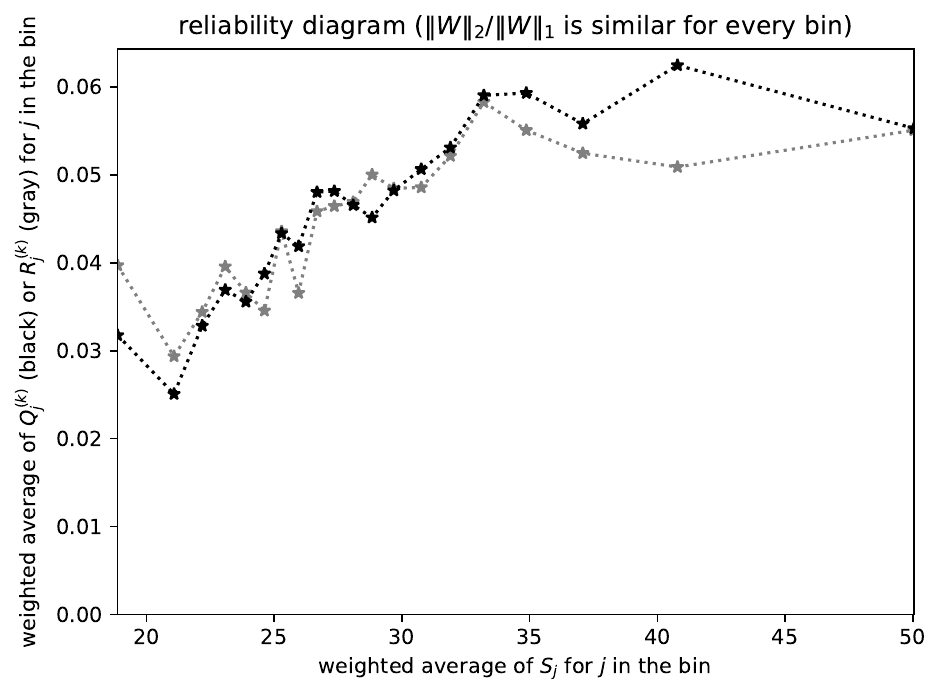}
}
\subfloat{
\includegraphics[width=\imsize]
{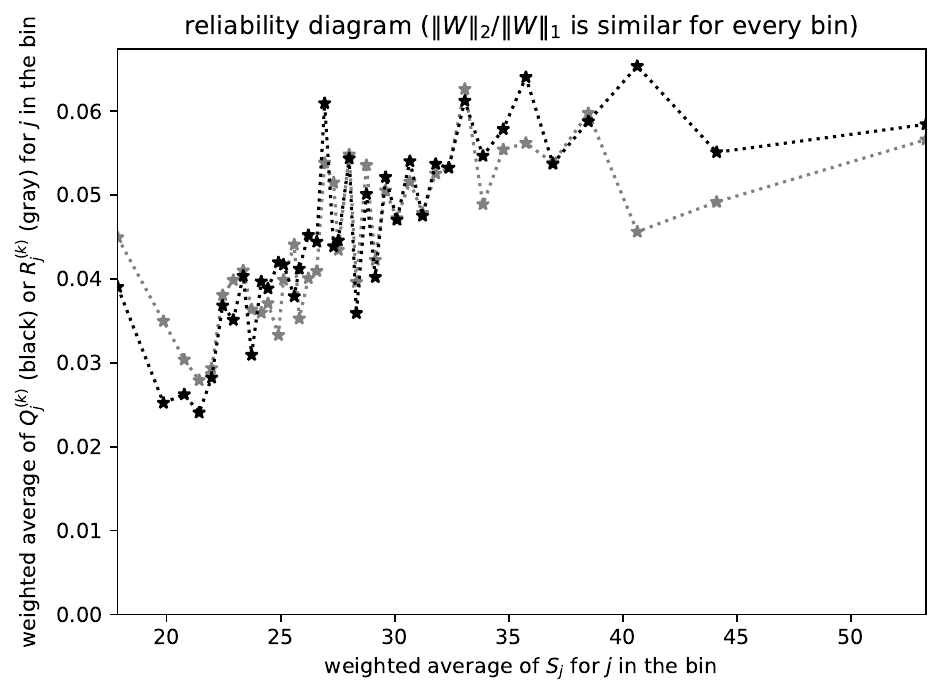}
}
\end{center}
\caption{
Have angina or coronary heart disease compared with having had a heart attack,
versus BMI. Discerning the range of lowest scores (that is, the BMIs)
over which the cumulative plot drops steeply is difficult
in all but the reliability diagram with the greatest number of bins.
And, unfortunately, the reliability diagram with the greatest number of bins
is very noisy. The reliability diagrams also look inconsistent around BMI
of 41 (the steep incline in the cumulative graph explains why).}
\label{angina_heart-attack}
\end{figure}

\begin{figure}
\begin{center}
\subfloat{
\includegraphics[width=\imsize]
{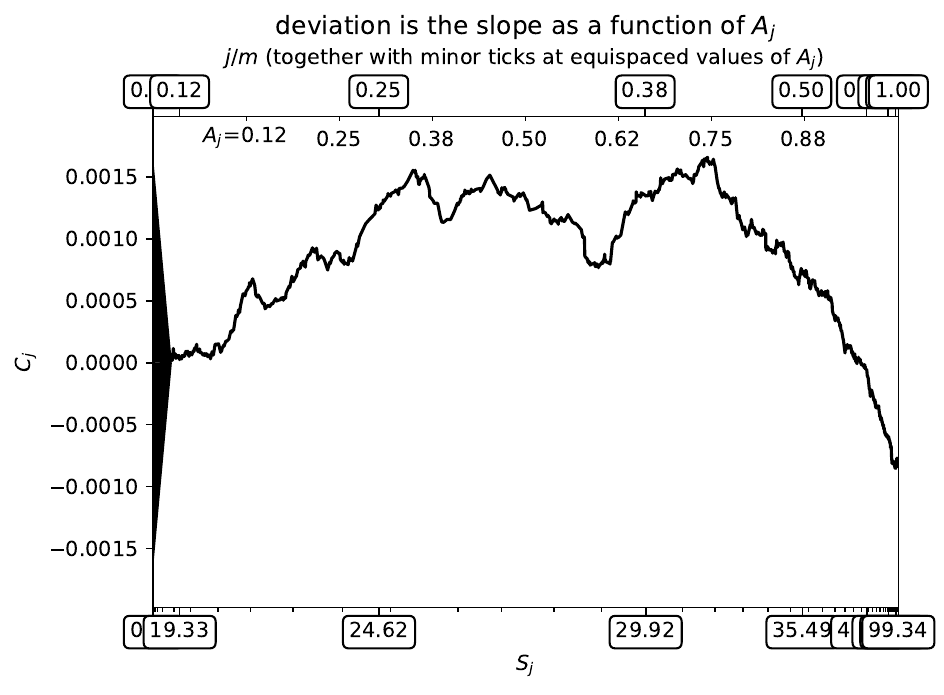}
}
\subfloat{
\parbox{\imsize}{\footnotesize
$m =$ 396,326 (with $\ell =$ 3,985 distinct scores) \\
$n =$ 3,985 \\
Kuiper's statistic $= 0.002511 / \sigma = 3.052$; the asymptotic P-value
$= 0.009106$ \\
Kolmogorov-Smirnov's $= 0.001659 / \sigma = 2.015$; asymptotic P-value
$= 0.08773$ \\
ATE $= -0.0008138 / \sigma = -0.9889$
\vspace{\fudger}
}
}

\vspace{-.7\fudger}

\subfloat{
\includegraphics[width=\imsize]
{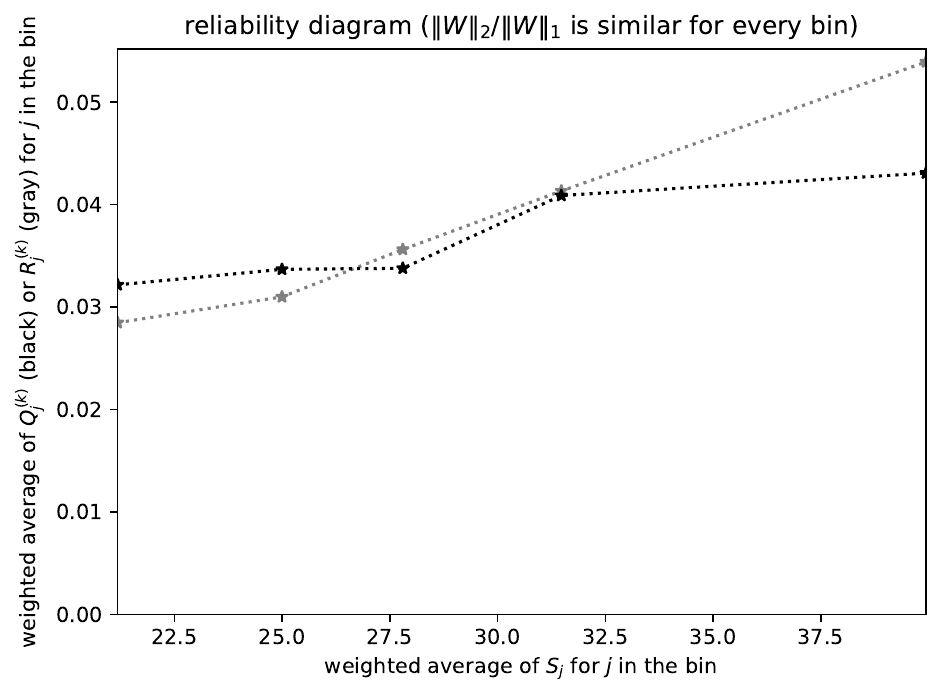}
}
\subfloat{
\includegraphics[width=\imsize]
{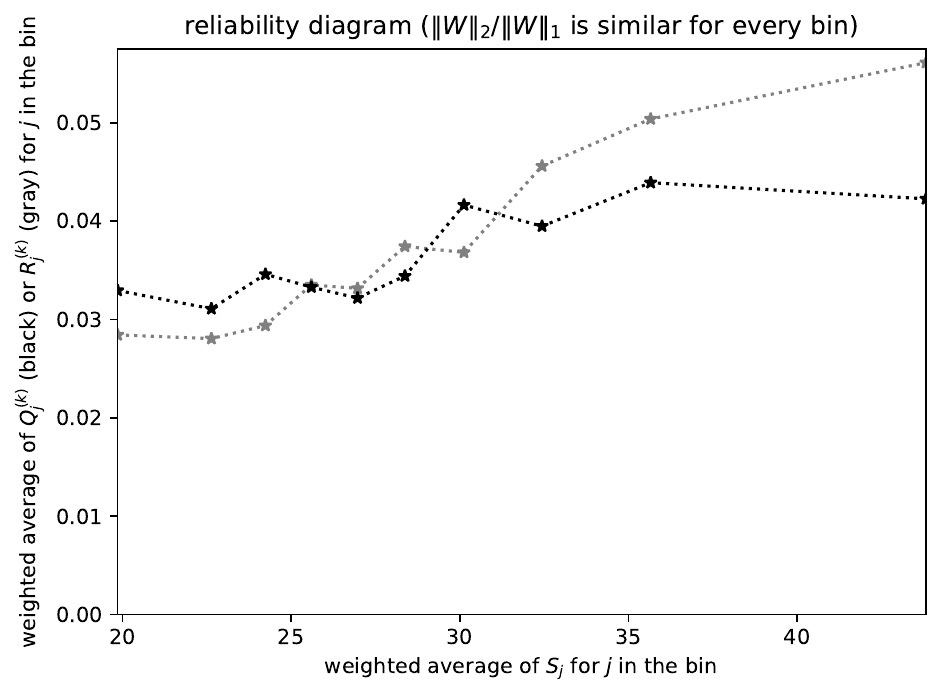}
}

\subfloat{
\includegraphics[width=\imsize]
{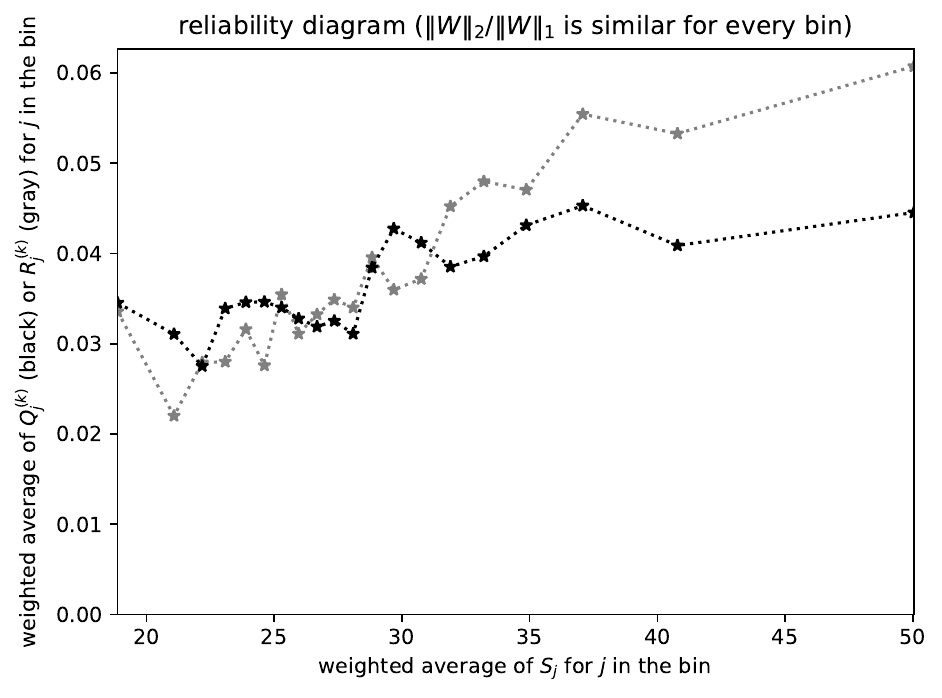}
}
\subfloat{
\includegraphics[width=\imsize]
{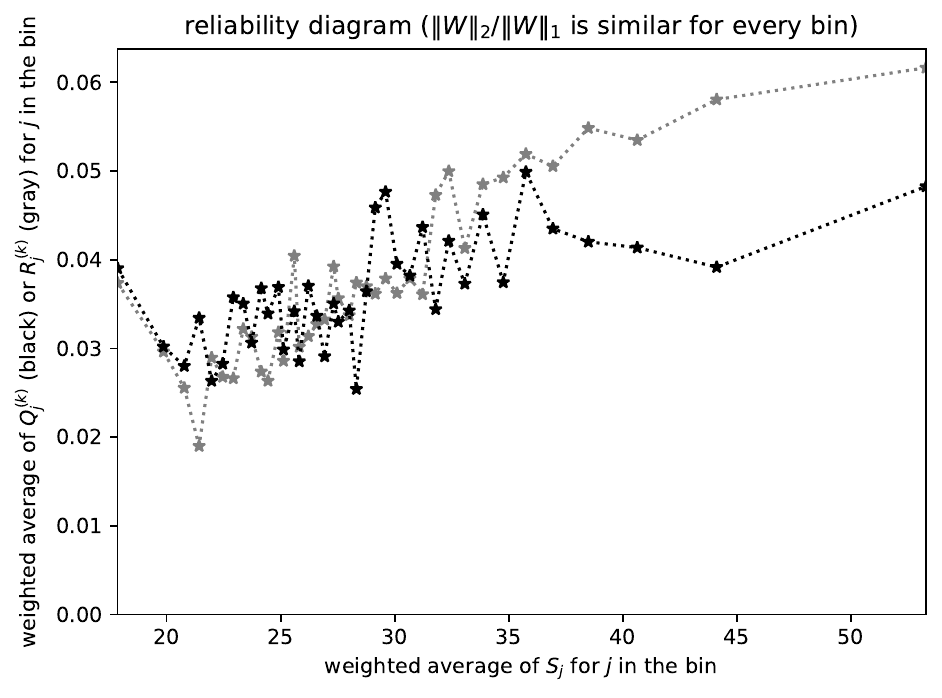}
}
\end{center}
\caption{
Have had a stroke compared with having kidney disease, versus BMI.
As with all Figures~\ref{angina_heart-attack}--\ref{hiv_kidney3},
the scores are the survey participants' BMIs.
Discerning that the expected difference between having had a stroke
and having kidney disease is roughly constant for scores greater than 35
is really hard from the reliability diagrams, due to noise
when there are many bins or insufficient resolution when there are few bins.}
\label{stroke_kidney-disease}
\end{figure}

\begin{figure}
\begin{center}
\subfloat{
\includegraphics[width=\imsize]
{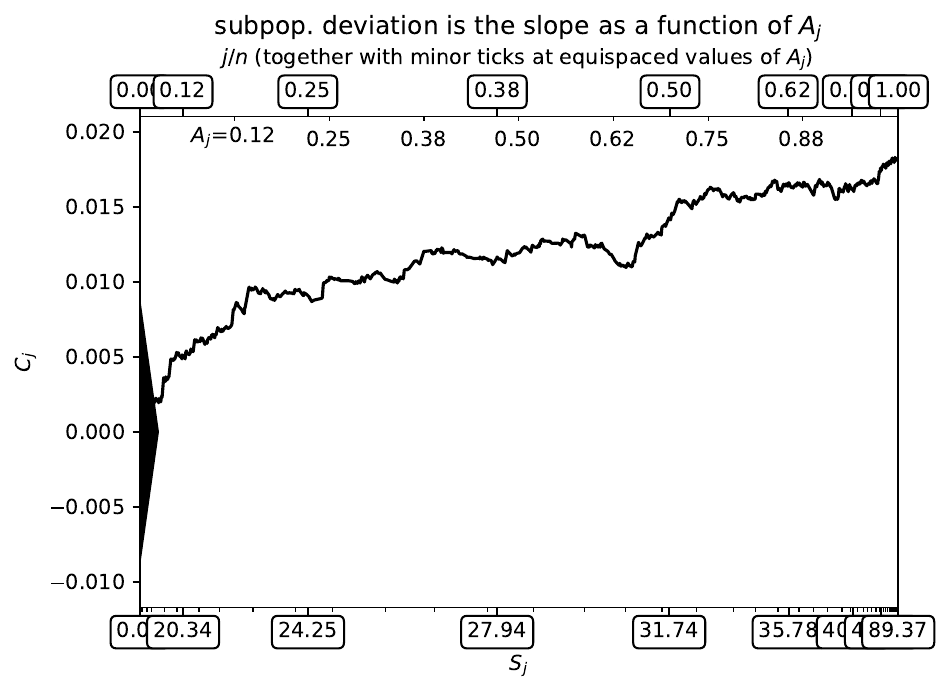}
}
\subfloat{
\parbox{\imsize}{\footnotesize
$m =$ 396,326 (with $\ell =$ 3,985 distinct scores) \\
$n_0 =$ 23,035 \\
$n =$ 3,985 \\
Kuiper's statistic $= 0.01831 / \sigma = 4.083$; the asymptotic P-value
$= 0.0001781$ \\
Kolmogorov-Smirnov's $= 0.01829 / \sigma = 4.078$; asymptotic P-value
$= 0.0000908$ \\
ATE $= 0.01821 / \sigma = 4.061$
\vspace{\fudger}
}
}

\vspace{-.7\fudger}

\subfloat{
\includegraphics[width=\imsize]
{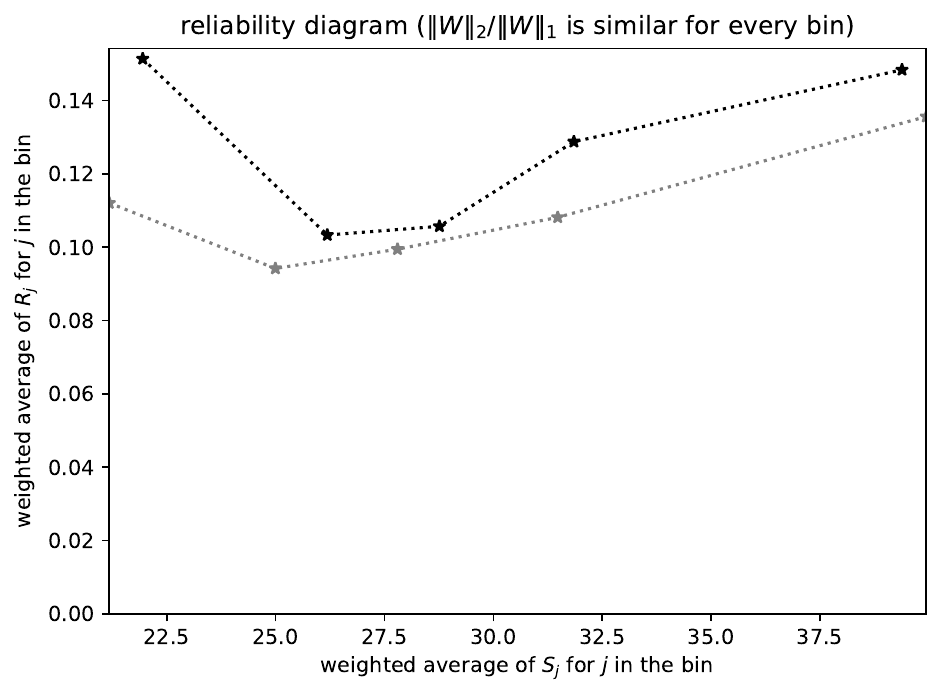}
}
\subfloat{
\includegraphics[width=\imsize]
{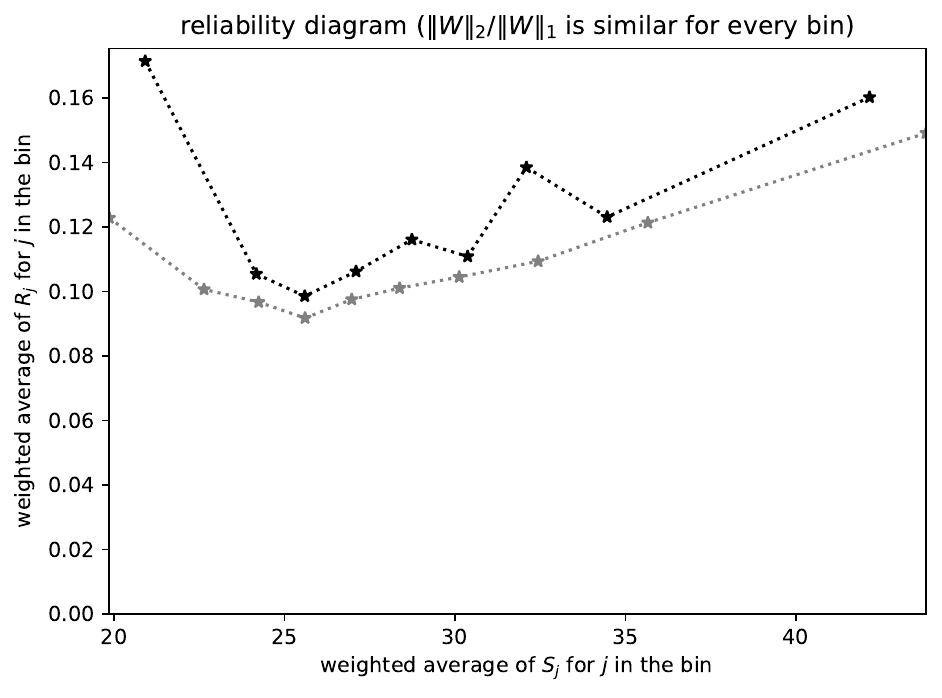}
}

\subfloat{
\includegraphics[width=\imsize]
{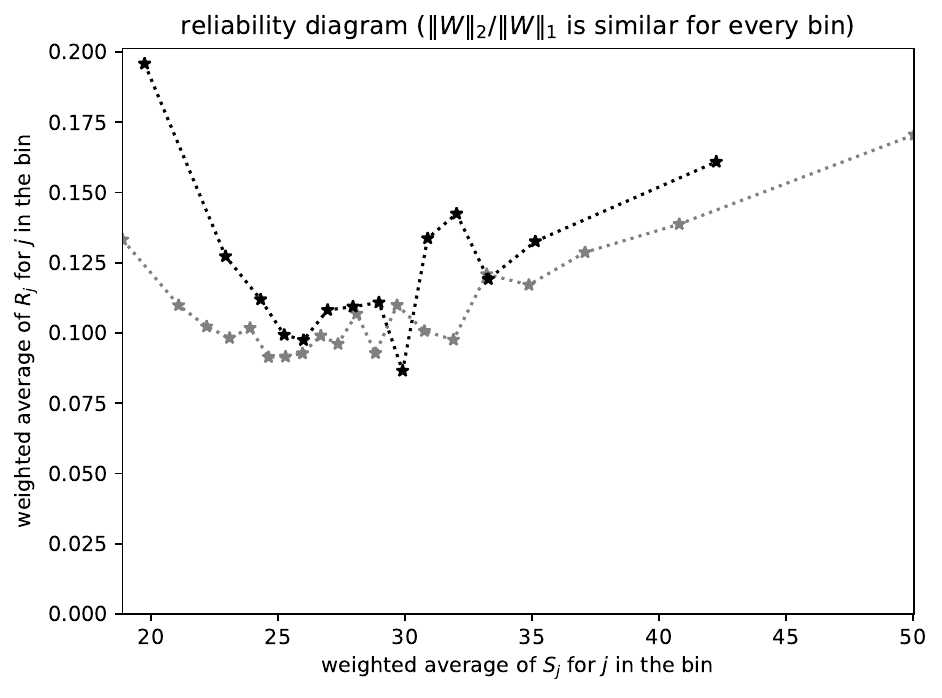}
}
\subfloat{
\includegraphics[width=\imsize]
{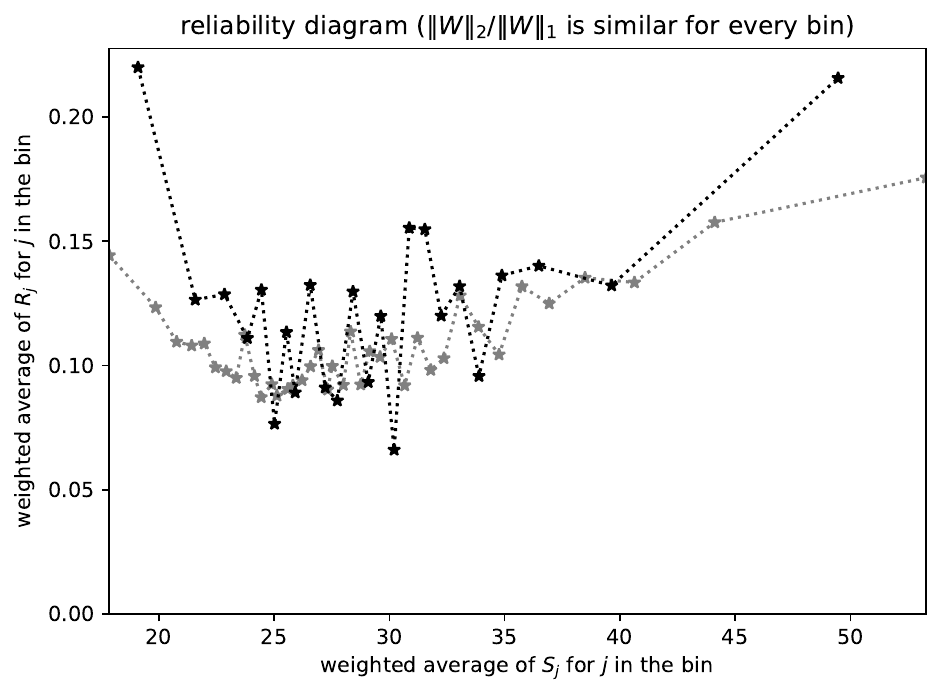}
}
\end{center}
\caption{
Could afford to see a doctor (rather than could not) versus BMI
for those who have had a heart attack compared with the full population
(the full population includes both those who have had a heart attack
and those who have not).
There seems to be a Simpson's Paradox in the reliability diagrams
for BMI around 30.
Determining the size of the difference for BMI less than 20
is impossible from any but the noisiest of the reliability diagrams,
whereas the difference is easy to quantify from the slope
of the cumulative graph.}
\label{heart-attack_doctor}
\end{figure}

\begin{figure}
\begin{center}
\subfloat{
\includegraphics[width=\imsize]
{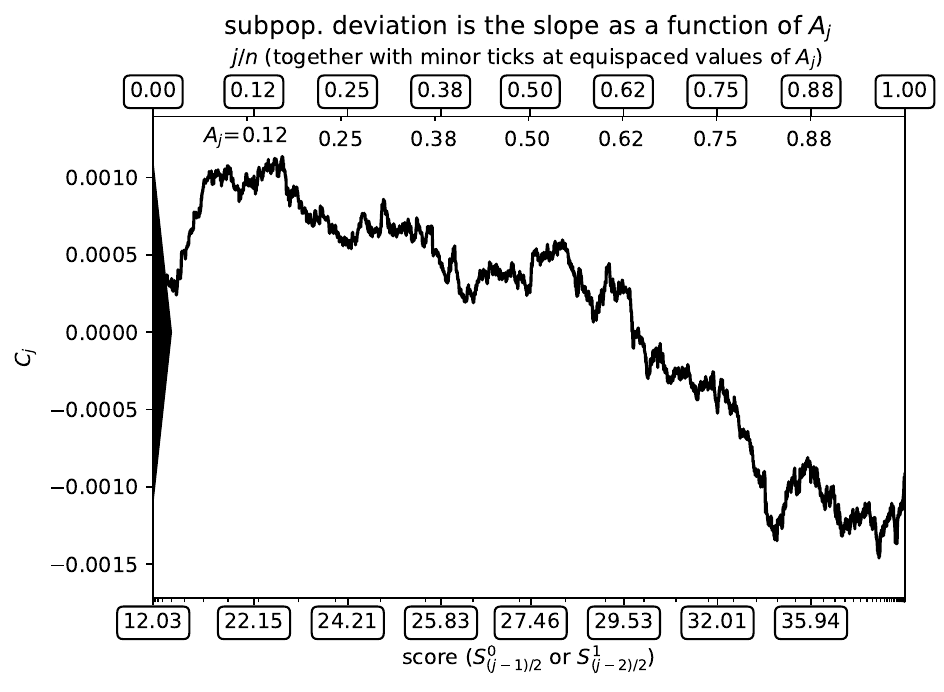}
}
\subfloat{
\parbox{\imsize}{\footnotesize
$m =$ 396,326 (with 3,985 distinct scores prior to randomization) \\
$n_0 =$ 121,203 \\
$n_1 =$ 232,734 \\
$n =$ 79,220 \\
Kuiper's statistic $= 0.002595 / \sigma = 4.635$; the asymptotic P-value
$= 0.000014$ \\
Kolmogorov-Smirnov's $= 0.001460 / \sigma = 2.607$; asymptotic P-value
$= 0.01827$ \\
ATE $= -0.001484 / \sigma = -2.650$ (or $-0.001711 / \sigma$ $= -3.056$
after having averaged over 25 random infinitesimal perturbations
of the original scores)
\vspace{\fudger}
}
}

\vspace{-.9\fudger}

\subfloat{
\includegraphics[width=\imsize]
{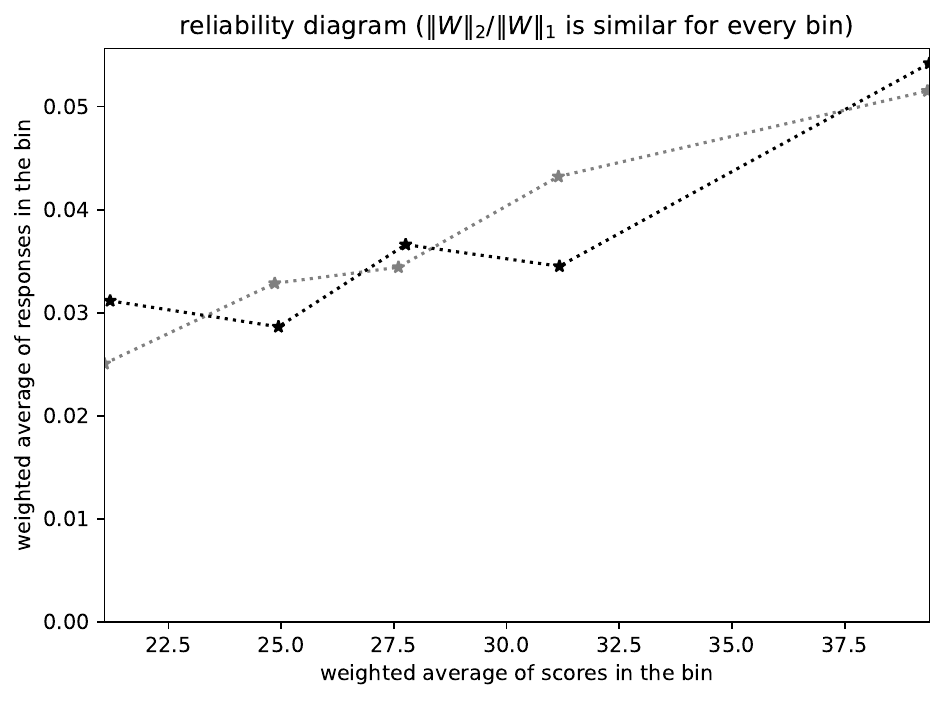}
}
\subfloat{
\includegraphics[width=\imsize]
{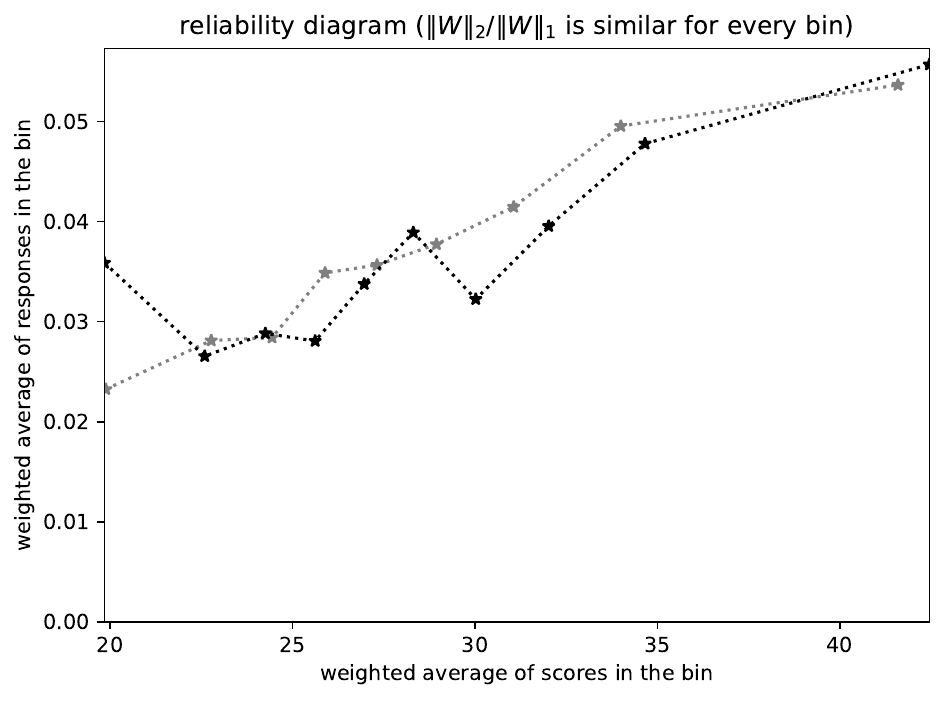}
}

\subfloat{
\includegraphics[width=\imsize]
{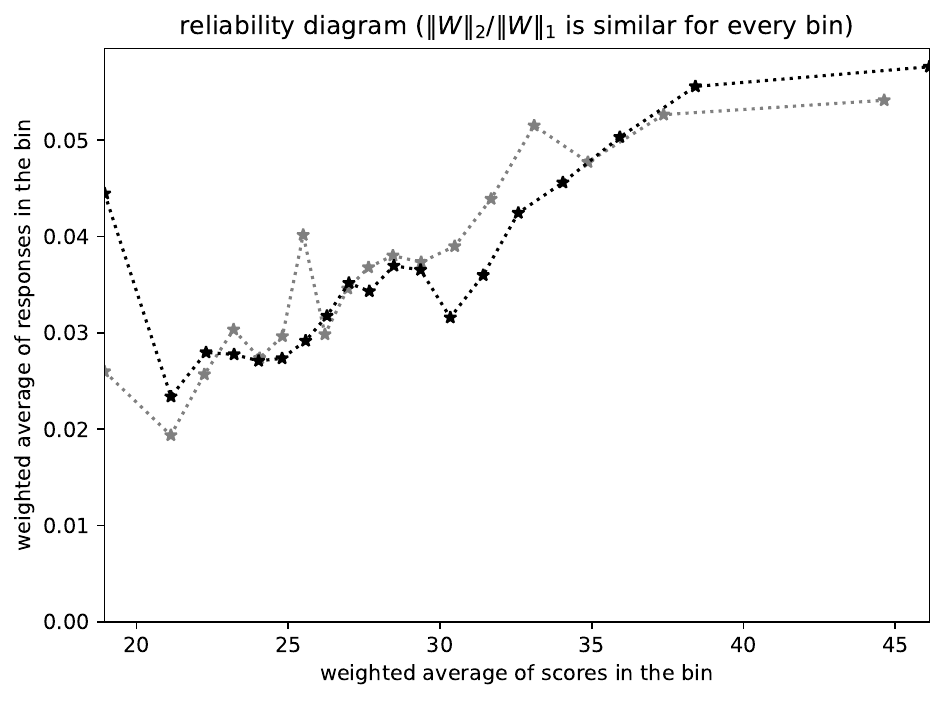}
}
\subfloat{
\includegraphics[width=\imsize]
{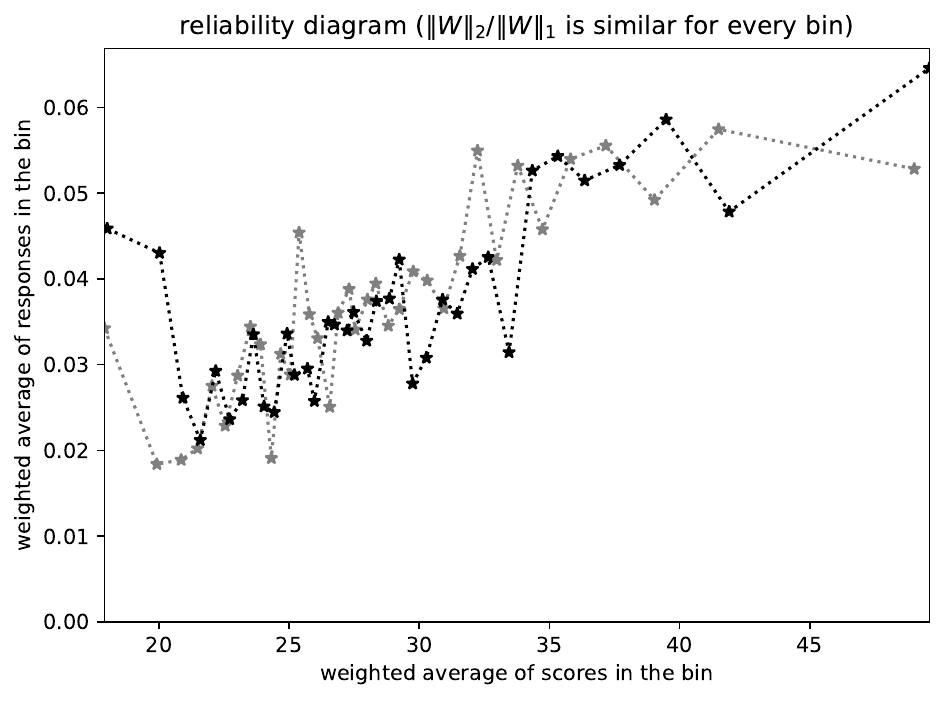}
}
\end{center}
\caption{
Have kidney disease (rather than do not) versus BMI
for those tested for HIV compared with those not tested.
Making sense for scores less than 20 of the large difference
between those tested and those not tested is very difficult
using only the reliability diagrams, whereas the cumulative plot
is crystal clear.}
\label{hiv_kidney1}
\end{figure}

\begin{figure}
\begin{center}
\subfloat{
\includegraphics[width=\imsize]
{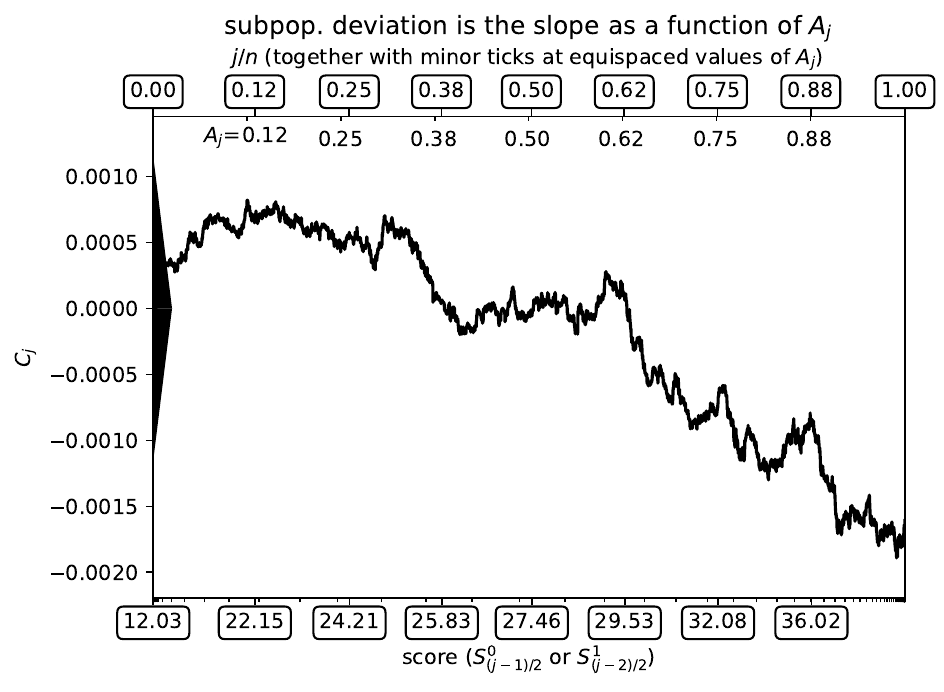}
}
\subfloat{
\parbox{\imsize}{\footnotesize
$m =$ 396,326 (with 3,985 distinct scores prior to randomization) \\
$n_0 =$ 121,203 \\
$n_1 =$ 232,734 \\
$n =$ 79,533 \\
Kuiper's statistic $= 0.002713 / \sigma = 4.713$; the asymptotic P-value
$= 0.0000097$ \\
Kolmogorov-Smirnov's $= 0.001892 / \sigma = 3.286$; asymptotic P-value
$= 0.002034$ \\
ATE $= -0.002477 / \sigma = -4.303$ (or $-0.001586 / \sigma$ $= -2.756$
after having averaged over 25 random infinitesimal perturbations
of the original scores)
\vspace{\fudger}
}
}

\vspace{-.9\fudger}

\subfloat{
\includegraphics[width=\imsize]
{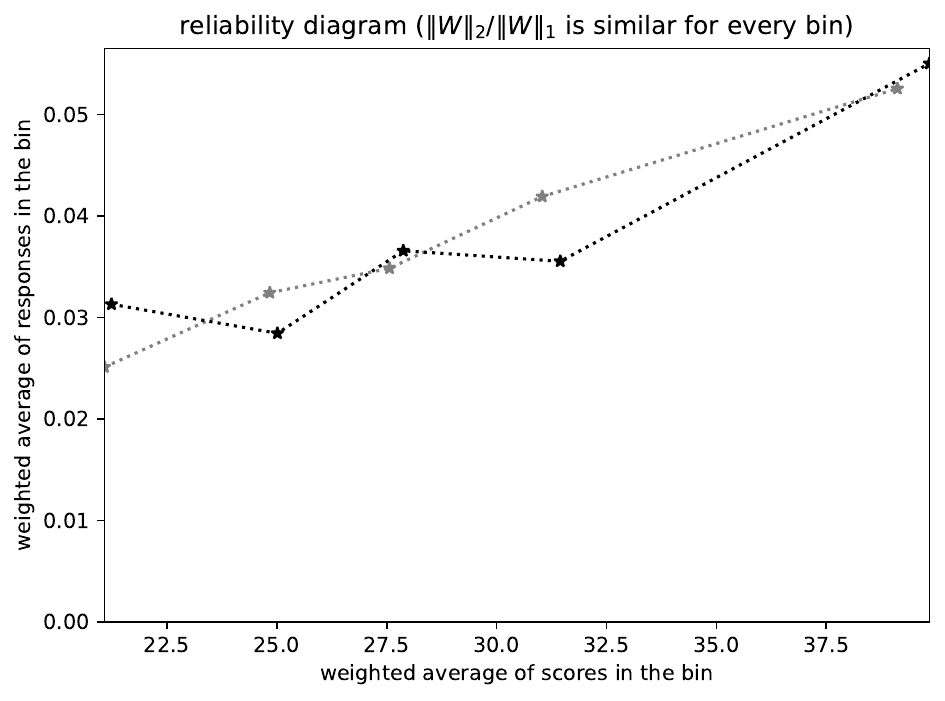}
}
\subfloat{
\includegraphics[width=\imsize]
{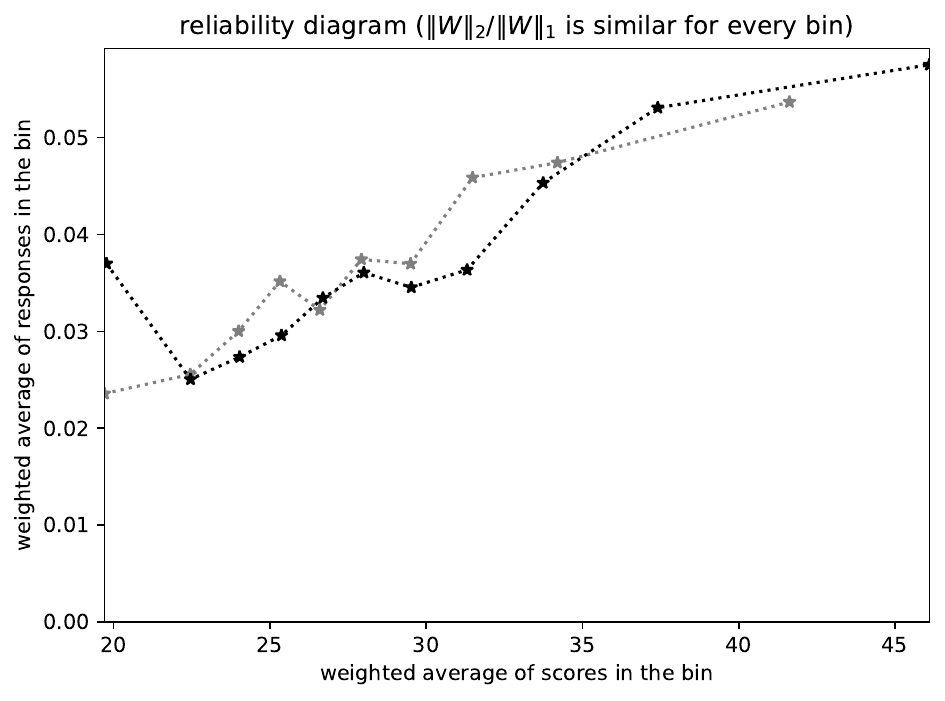}
}

\subfloat{
\includegraphics[width=\imsize]
{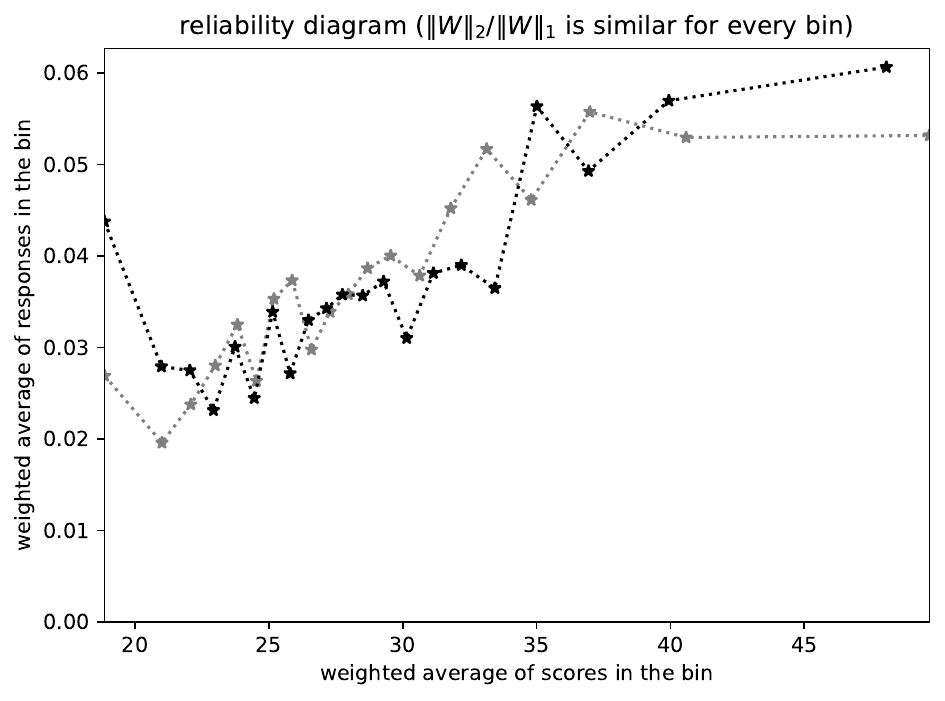}
}
\subfloat{
\includegraphics[width=\imsize]
{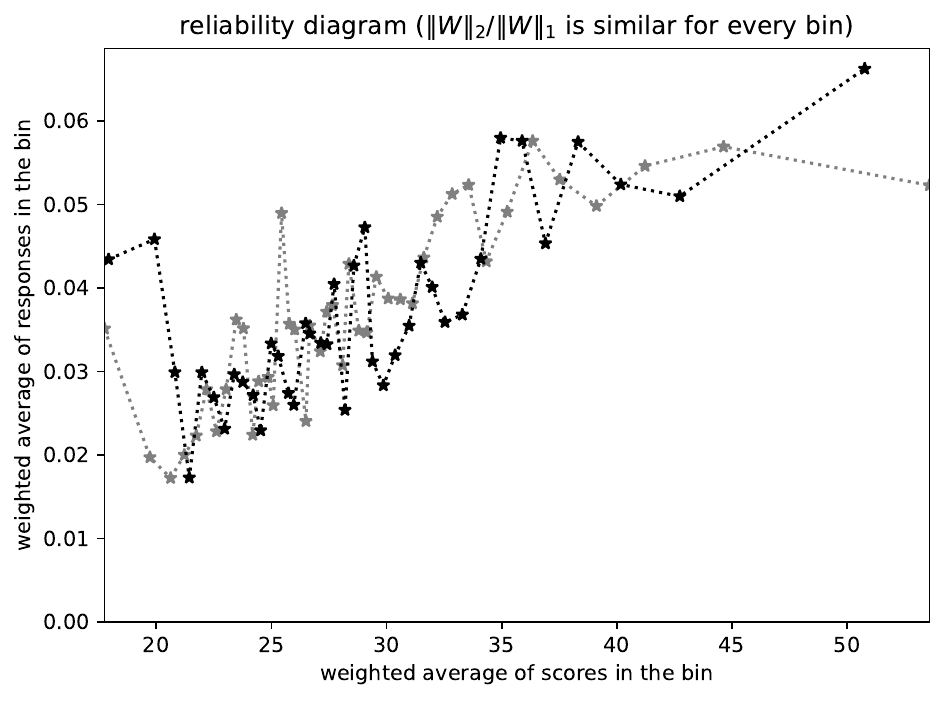}
}
\end{center}
\caption{
Have kidney disease (rather than do not) versus BMI
for those tested for HIV compared with those not tested, with scores randomized
from a different random seed (namely, 54321) than that
for Figure~\ref{hiv_kidney1} (which used 543216789).
As in Figure~\ref{hiv_kidney1}, making sense for scores below 20
of the big difference between those tested and those not tested
is hard using only the reliability diagrams,
whereas the cumulative plots are clear.}
\label{hiv_kidney2}
\end{figure}

\begin{figure}
\begin{center}
\subfloat{
\includegraphics[width=\imsize]
{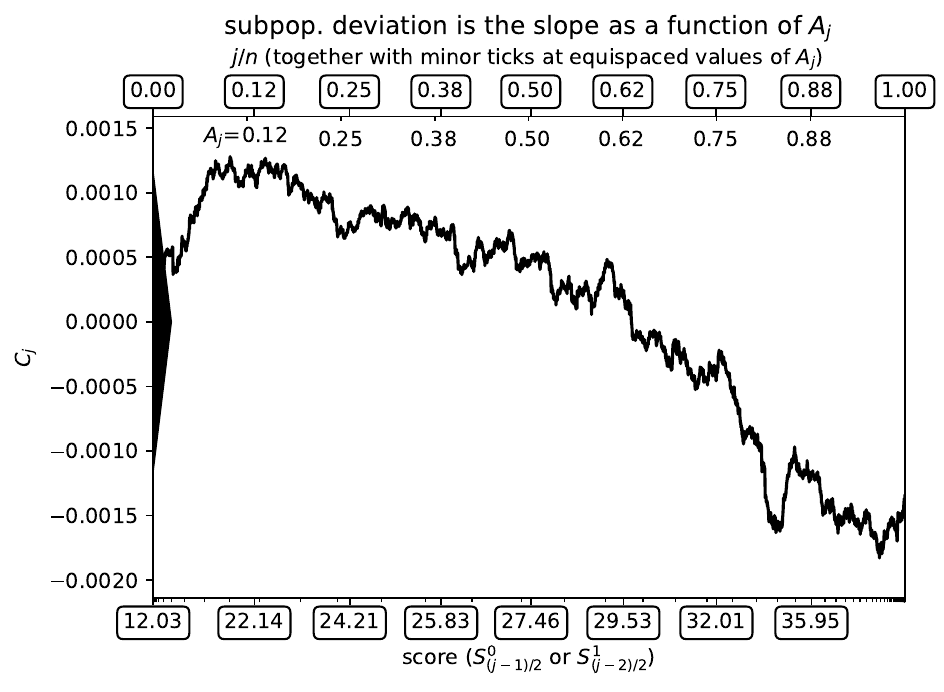}
}
\subfloat{
\parbox{\imsize}{\footnotesize
$m =$ 396,326 (with 3,985 distinct scores prior to randomization) \\
$n_0 =$ 121,203 \\
$n_1 =$ 232,734 \\
$n =$ 79,551 \\
Kuiper's statistic $= 0.003108 / \sigma = 5.214$; the asymptotic P-value
$= 0.0000007$ \\
Kolmogorov-Smirnov's $= 0.001828 / \sigma = 3.066$; asymptotic P-value
$= 0.004335$ \\
ATE $= -0.001858 / \sigma = -3.117$ (or $-0.001662 / \sigma$ $= -2.788$
after having averaged over 25 random infinitesimal perturbations
of the original scores)
\vspace{\fudger}
}
}

\vspace{-.9\fudger}

\subfloat{
\includegraphics[width=\imsize]
{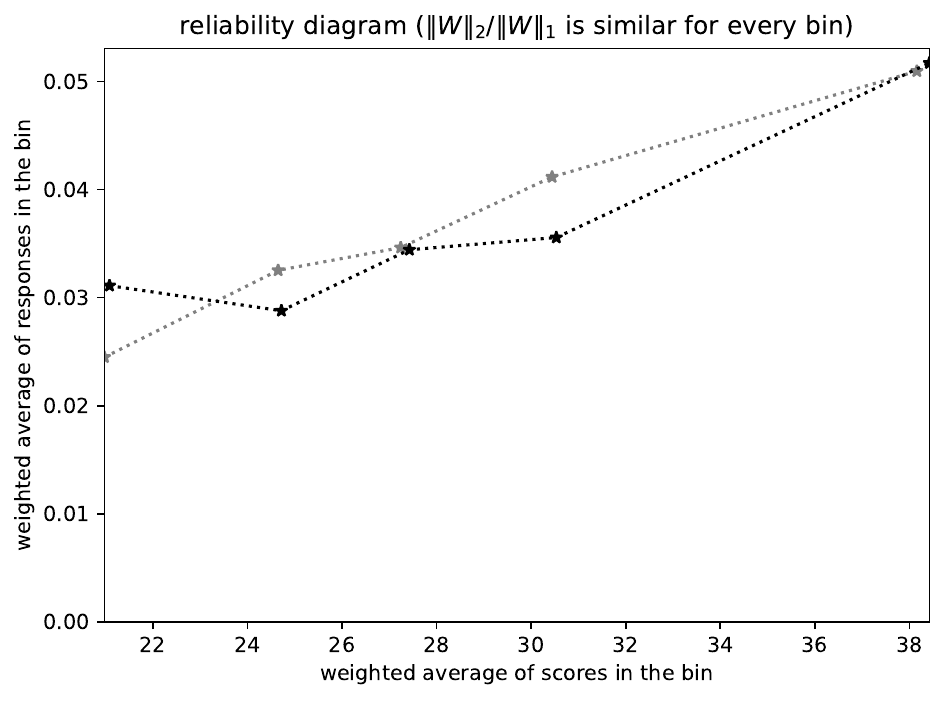}
}
\subfloat{
\includegraphics[width=\imsize]
{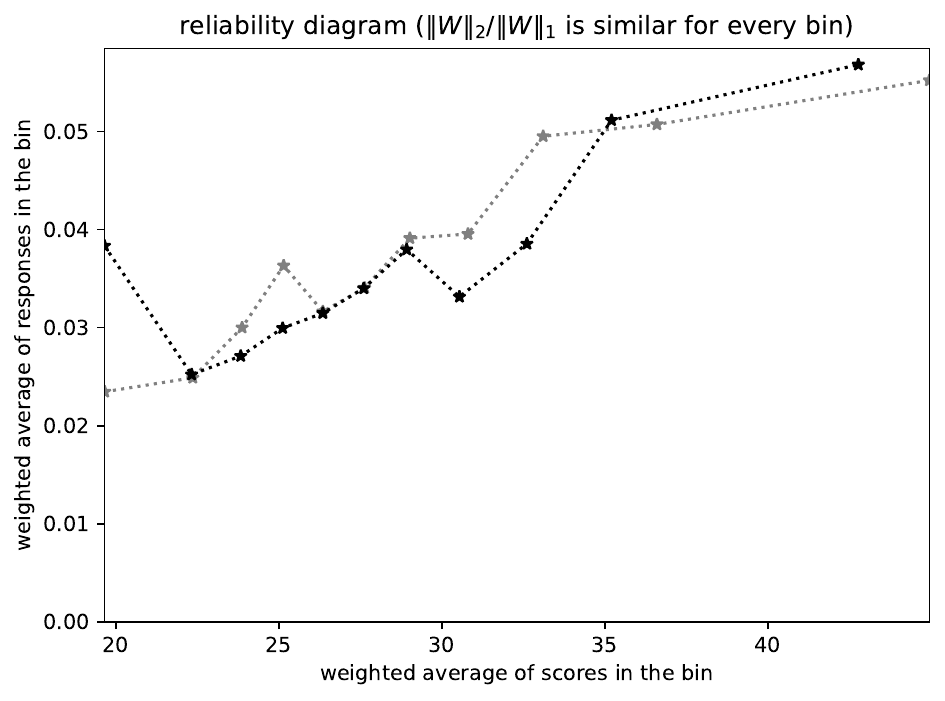}
}

\subfloat{
\includegraphics[width=\imsize]
{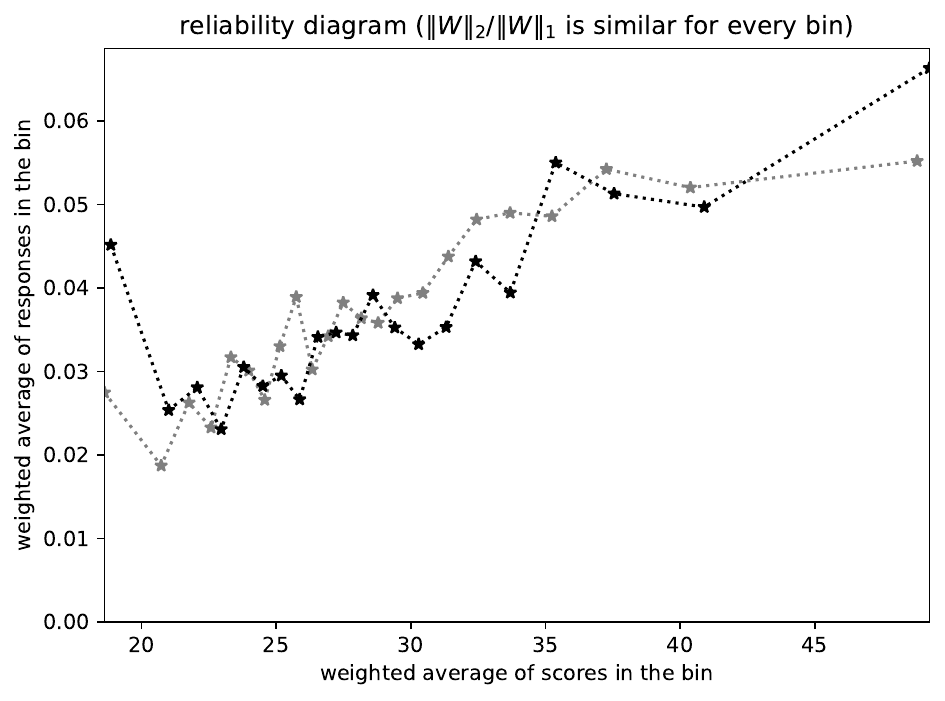}
}
\subfloat{
\includegraphics[width=\imsize]
{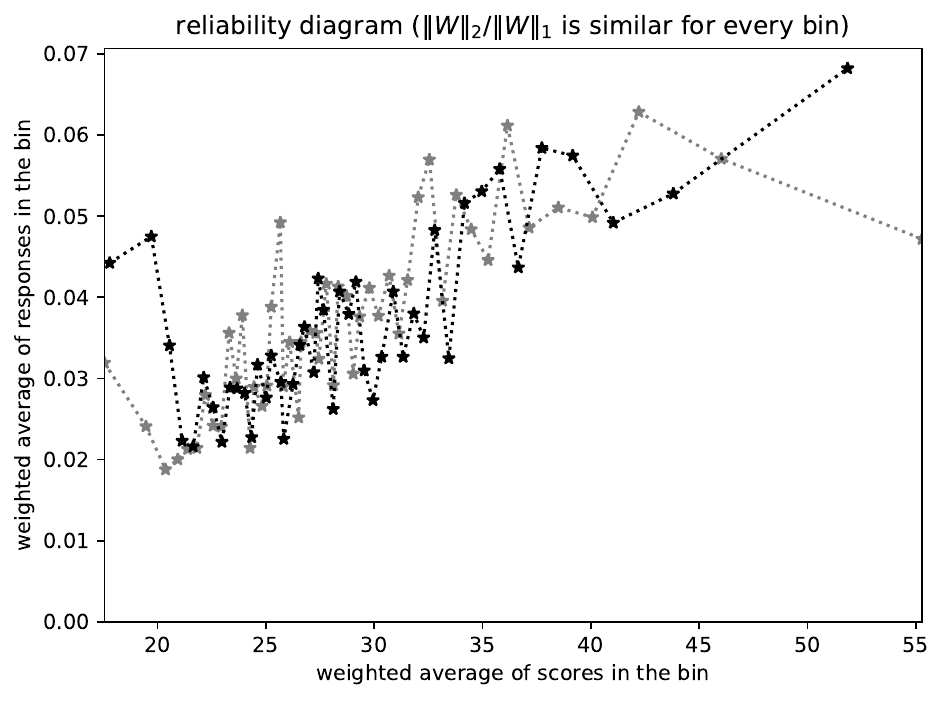}
}
\end{center}
\caption{
Have kidney disease (rather than do not) versus BMI
for those tested for HIV compared with those not tested, with scores randomized
from a different random seed (namely, 6789) than that
for Figures~\ref{hiv_kidney1} and~\ref{hiv_kidney2}.
The cumulative plot here is as clear as in Figures~\ref{hiv_kidney1}
and~\ref{hiv_kidney2}, whereas making sense for scores less than 20
of the large difference between those tested and those not tested
is hard using only the reliability diagrams.}
\label{hiv_kidney3}
\end{figure}

\begin{figure}
\begin{center}
\subfloat{
\includegraphics[width=\imsize]
{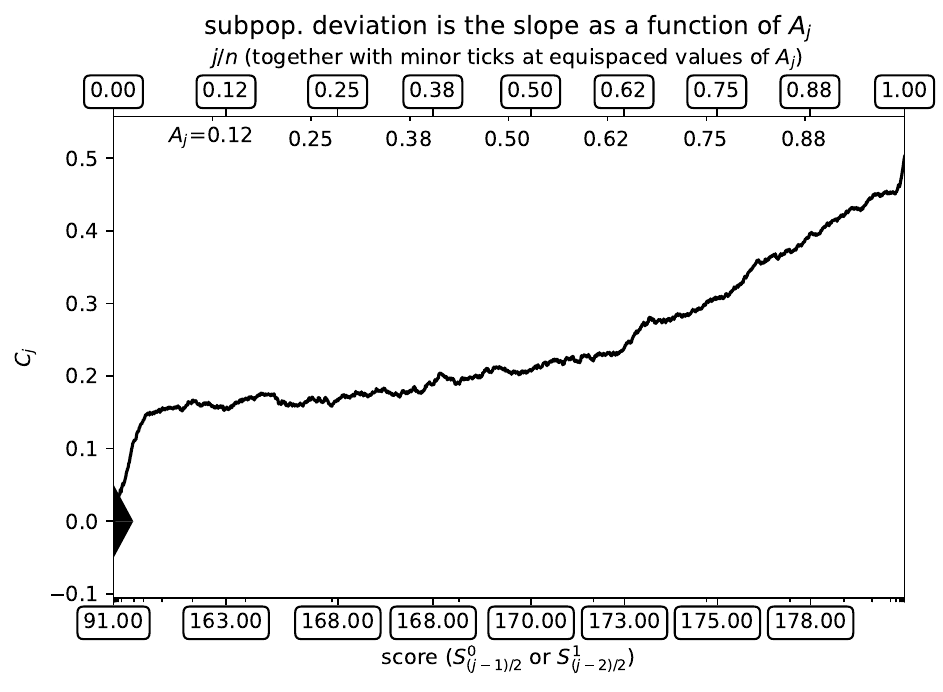}
}
\subfloat{
\parbox{\imsize}{\footnotesize
$m =$ 396,326 (with 105 distinct scores prior to randomization) \\
$n_0 =$ 193,659 \\
$n_1 =$ 202,667 \\
$n =$ 45,370 \\
Kuiper's statistic $= 0.5027 / \sigma = 19.88$;
the asymptotic P-value is less than $10^{-16}$ \\
Kolmogorov-Smirnov's $= 0.5021 / \sigma = 19.86$;
the asymptotic P-value is less than $10^{-16}$ \\
ATE $= 0.9044 / \sigma = 35.77$ (or $0.8513 / \sigma = 33.67$
following averaging over 25 random infinitesimal perturbations
of the original scores)
\vspace{\fudger}
}
}

\vspace{-\fudger}

\subfloat{
\includegraphics[width=\imsize]
{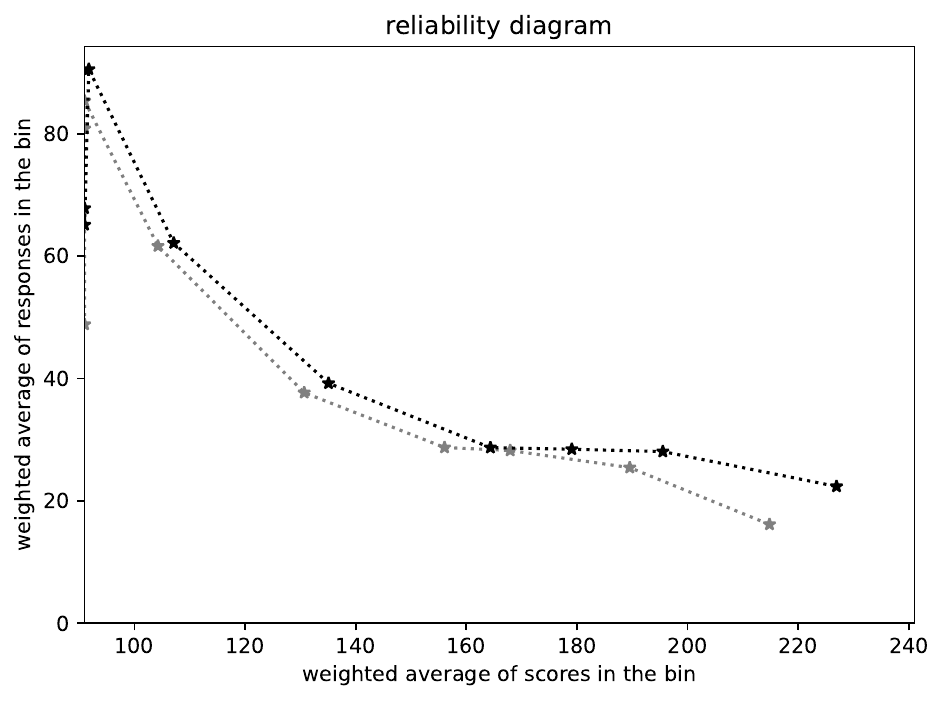}
}
\subfloat{
\includegraphics[width=\imsize]
{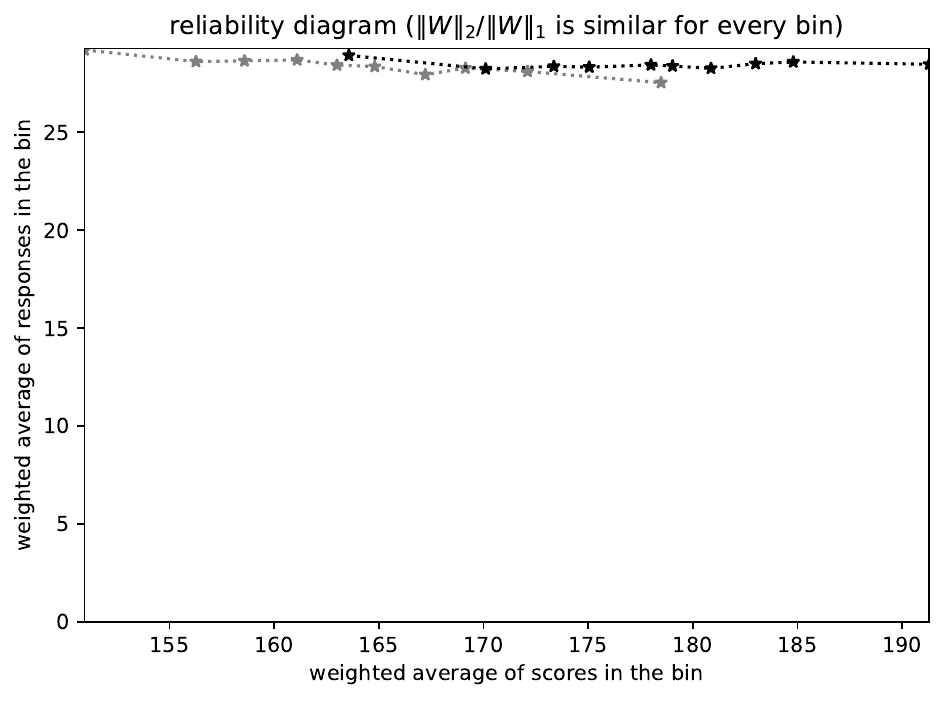}
}

\subfloat{
\includegraphics[width=\imsize]
{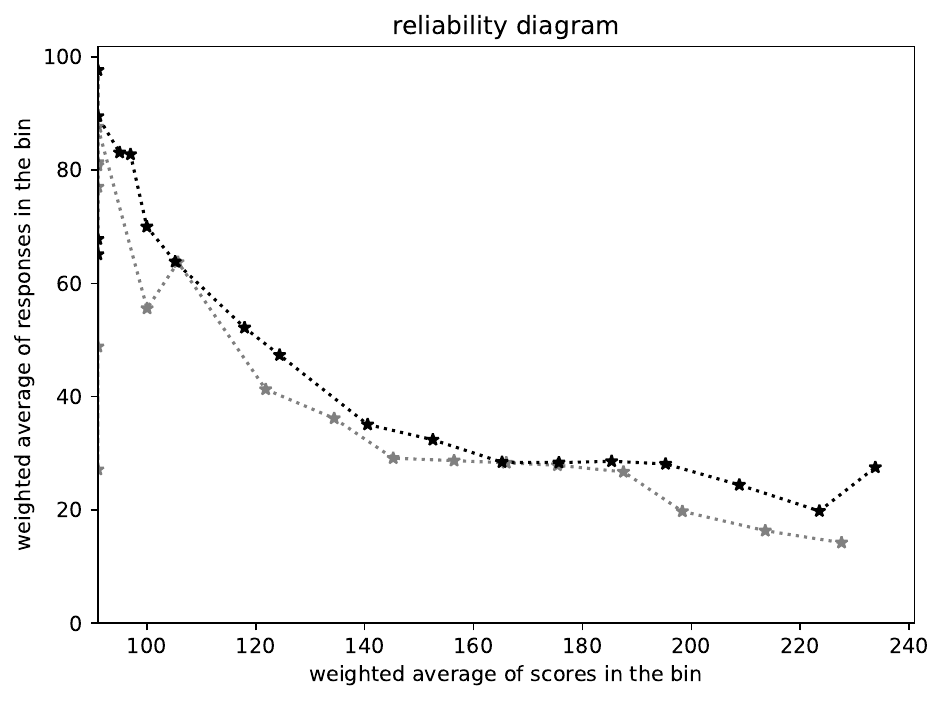}
}
\subfloat{
\includegraphics[width=\imsize]
{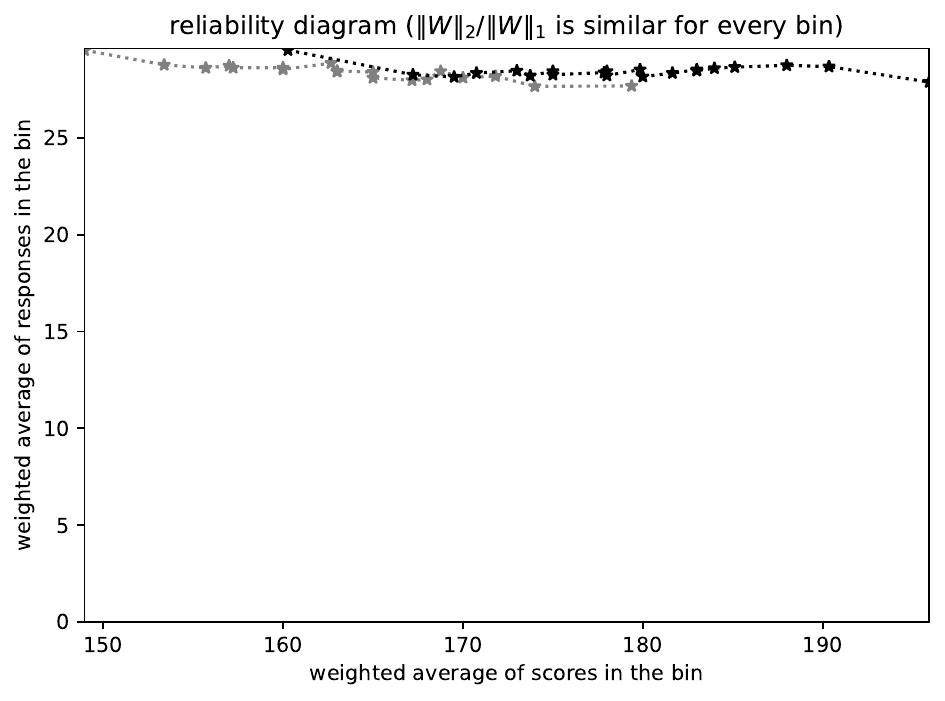}
}

\subfloat{
\includegraphics[width=\imsize]
{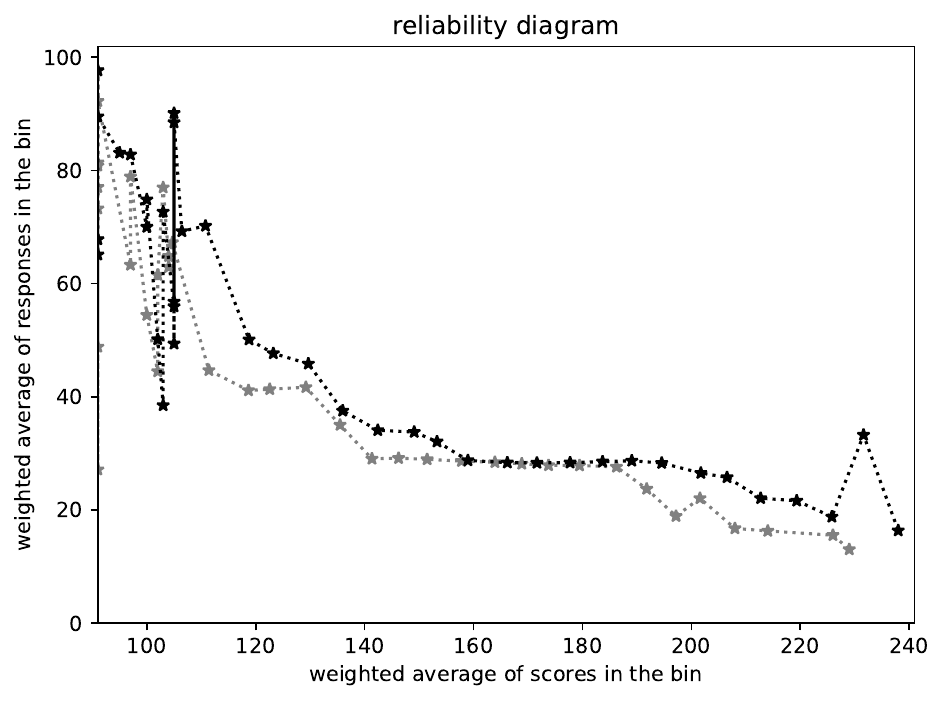}
}
\subfloat{
\includegraphics[width=\imsize]
{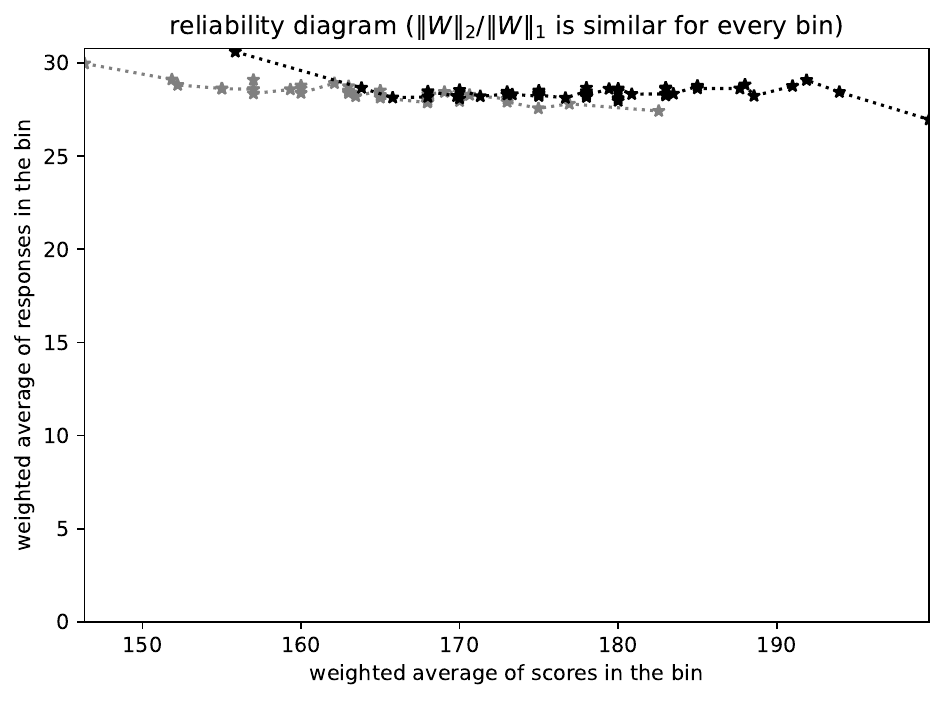}
}
\end{center}
\caption{BMI versus height in centimeters for men compared with women.
The scores for Figures~\ref{men_women1}--\ref{men_women3} are heights
instead of the BMIs used as scores for the earlier figures;
BMIs are still used here, but now as responses rather than scores.
Of course, height need not ``cause'' the associated BMI,
but a causal connection seems more plausible with BMI depending on height
rather than BMI ``causing'' height.
Much higher BMI for men than for women who are equally very short
jumps out of the cumulative plots. Assessing how much higher is trivial
from the slope of the cumulative graph yet very tricky to divine
from the reliability diagrams.}
\label{men_women1}
\end{figure}

\begin{figure}
\begin{center}
\subfloat{
\includegraphics[width=\imsize]
{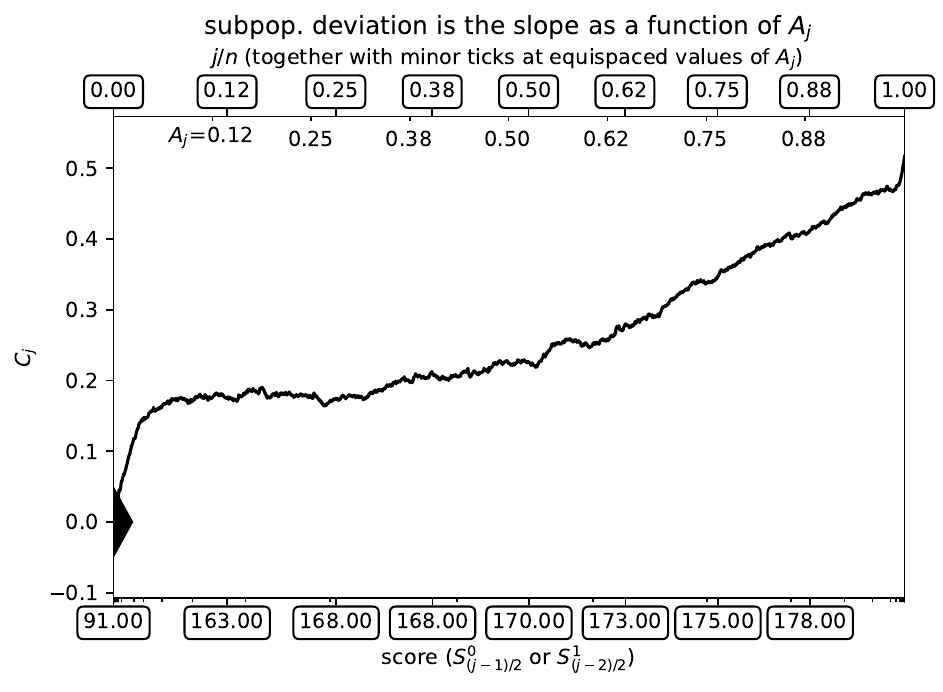}
}
\subfloat{
\parbox{\imsize}{\footnotesize
$m =$ 396,326 (with 105 distinct scores prior to randomization) \\
$n_0 =$ 193,659 \\
$n_1 =$ 202,667 \\
$n =$ 45,489 \\
Kuiper's statistic $= 0.5165 / \sigma = 20.29$;
the asymptotic P-value is less than $10^{-16}$ \\
Kolmogorov-Smirnov's $= 0.5165 / \sigma = 20.29$;
the asymptotic P-value is less than $10^{-16}$ \\
ATE $= 0.8183 / \sigma = 32.15$ (or $0.8603 / \sigma = 33.80$
following averaging over 25 random infinitesimal perturbations
of the original scores)
\vspace{\fudger}
}
}

\vspace{-\fudger}

\subfloat{
\includegraphics[width=\imsize]
{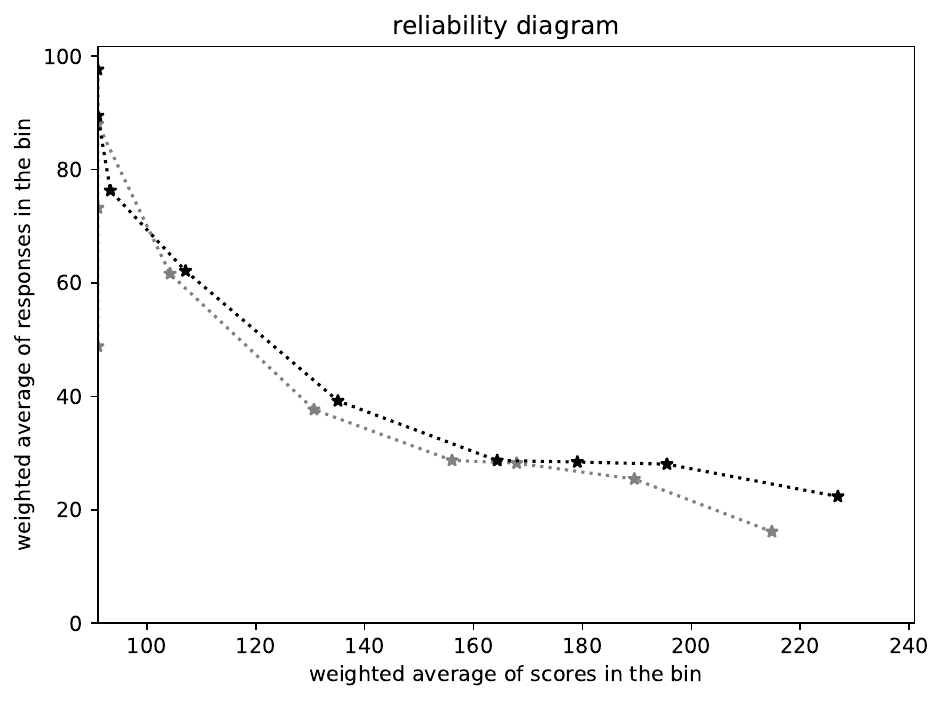}
}
\subfloat{
\includegraphics[width=\imsize]
{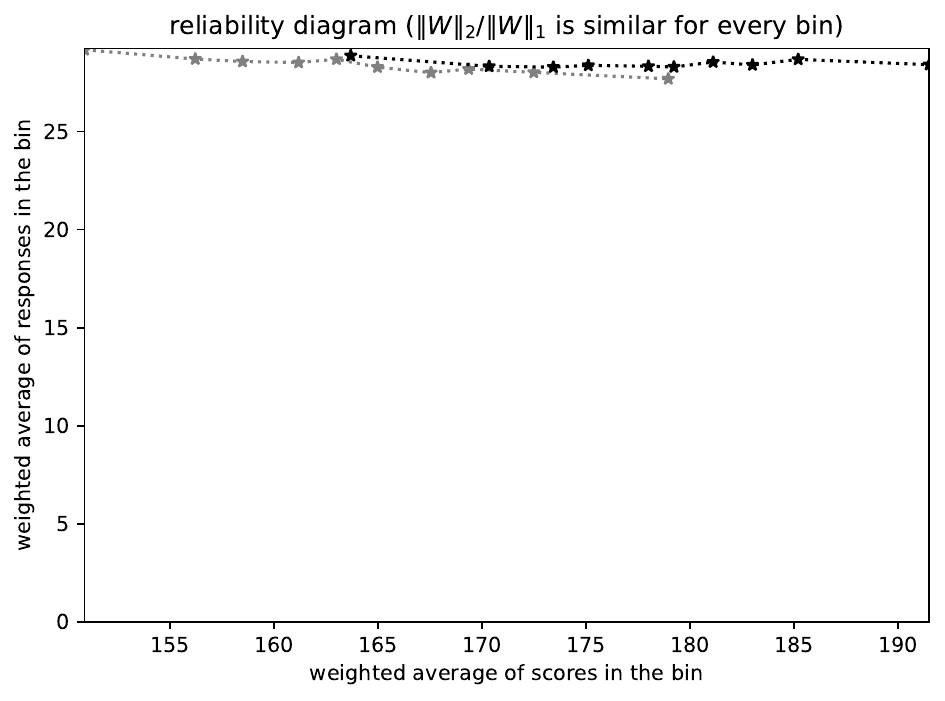}
}

\subfloat{
\includegraphics[width=\imsize]
{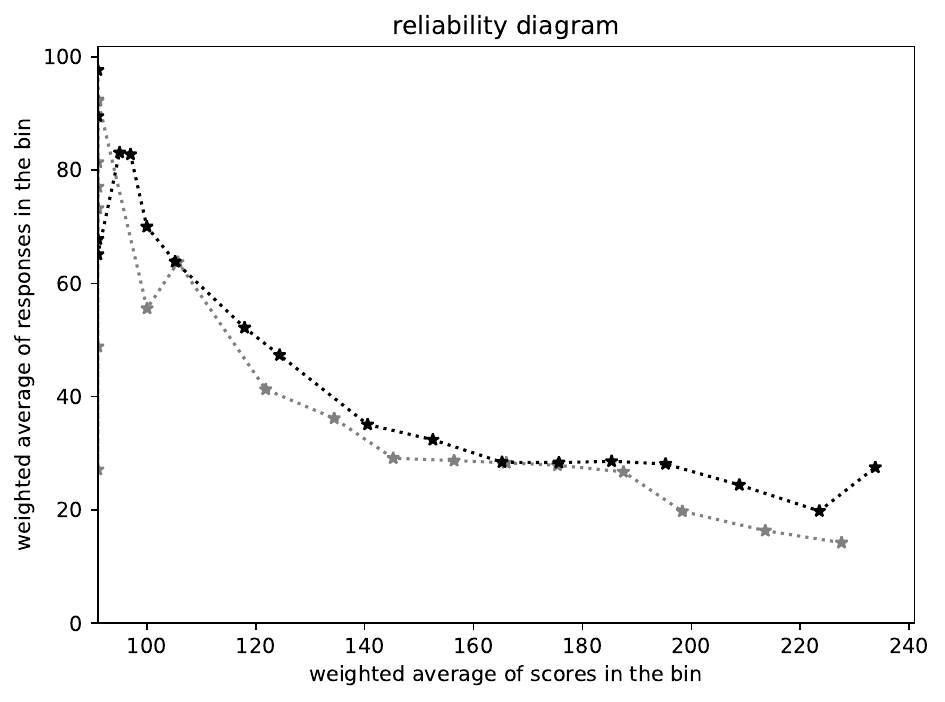}
}
\subfloat{
\includegraphics[width=\imsize]
{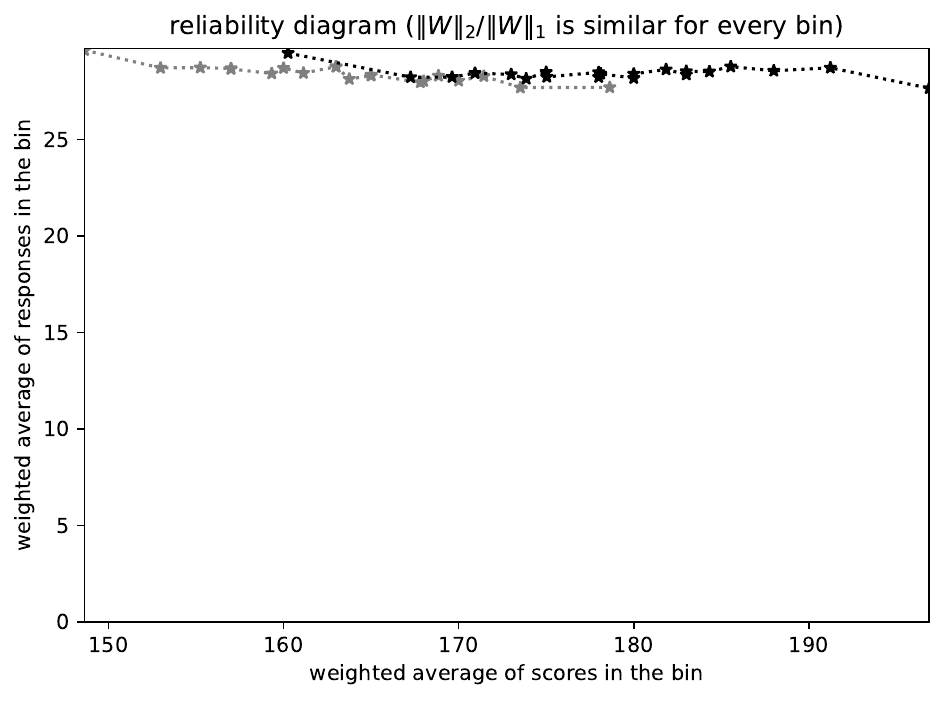}
}

\subfloat{
\includegraphics[width=\imsize]
{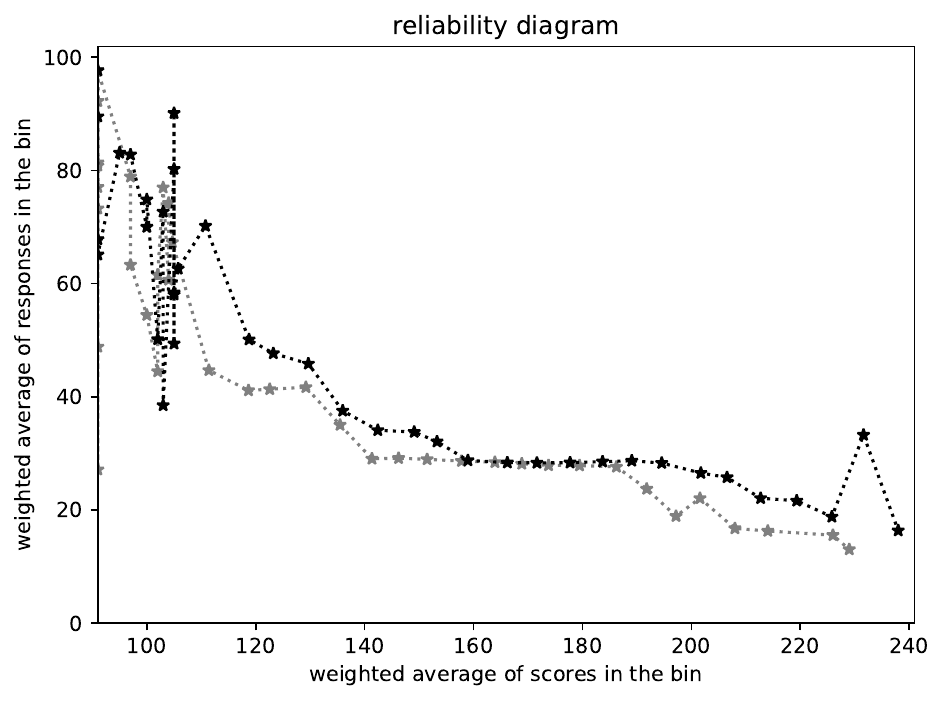}
}
\subfloat{
\includegraphics[width=\imsize]
{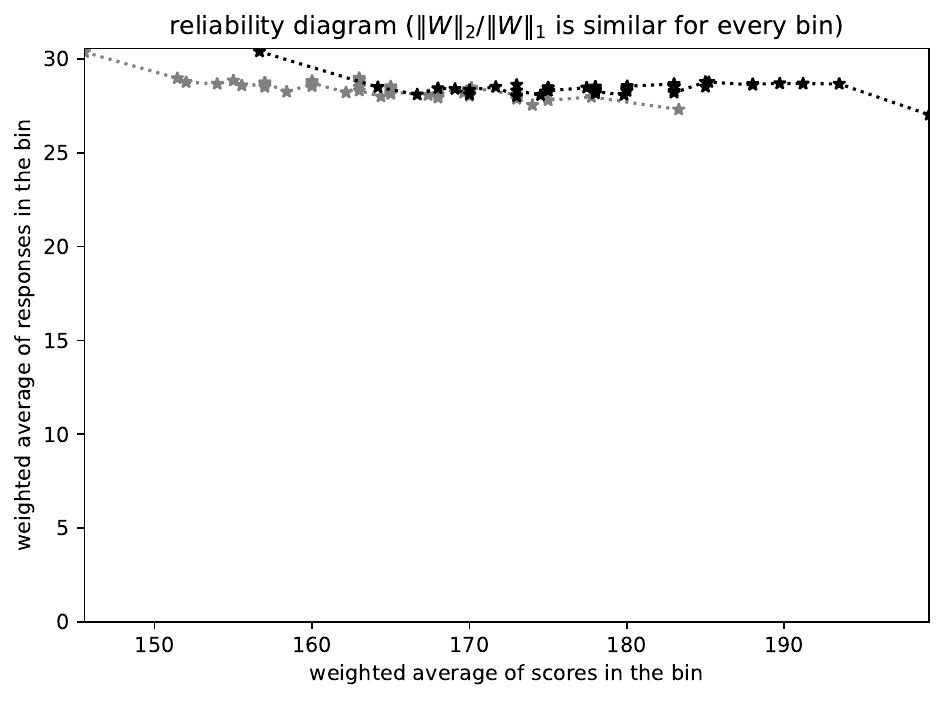}
}
\end{center}
\caption{BMI versus height in centimeters for men compared with women,
with scores perturbed at random starting with a random seed, 54321,
that is different from the seed used in Figure~\ref{men_women1}
(which was 543216789).
The scores for these figures are heights, instead of the BMIs used as scores
for the earlier figures; here, BMIs are responses rather than scores.
As in Figure~\ref{men_women1}, the cumulative plot readily reveals much higher
BMI for men than for women who report to be extremely short.
The slope of the cumulative graph quantifies how much higher,
whereas the reliability diagrams are hard to interpret
for the very small scores.}
\label{men_women2}
\end{figure}

\begin{figure}
\begin{center}
\subfloat{
\includegraphics[width=\imsize]
{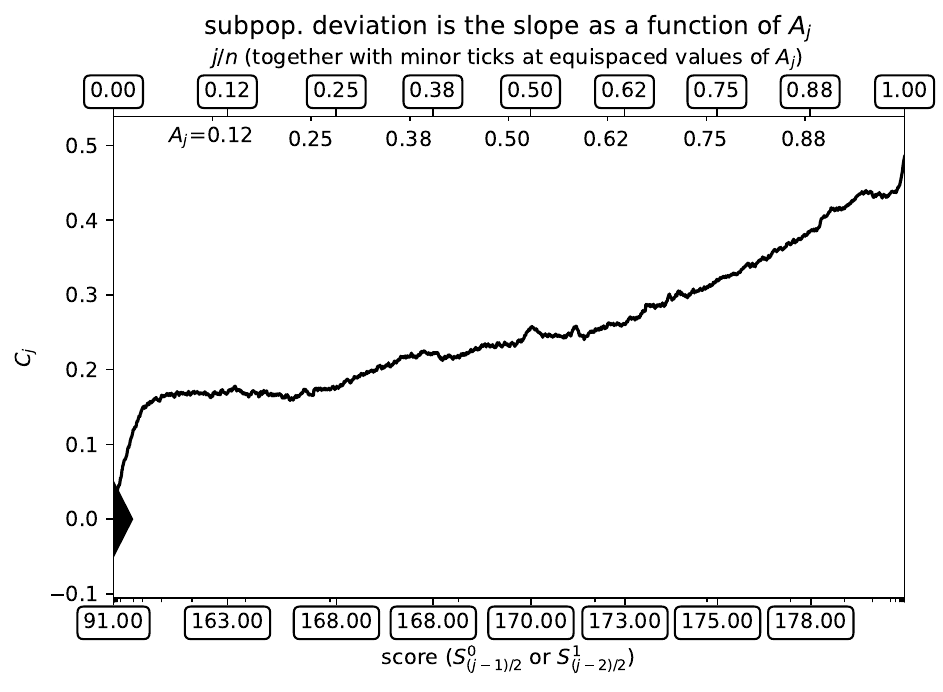}
}
\subfloat{
\parbox{\imsize}{\footnotesize
$m =$ 396,326 (with 105 distinct scores prior to randomization) \\
$n_0 =$ 193,659 \\
$n_1 =$ 202,667 \\
$n =$ 45,379 \\
Kuiper's statistic $= 0.4851 / \sigma = 18.62$;
the asymptotic P-value is less than $10^{-16}$ \\
Kolmogorov-Smirnov's $= 0.4851 / \sigma = 18.62$;
the asymptotic P-value is less than $10^{-16}$ \\
ATE $= 0.8651 / \sigma = 33.21$ (or $0.8432 / \sigma = 32.36$
following averaging over 25 random infinitesimal perturbations
of the original scores)
\vspace{\fudger}
}
}

\vspace{-\fudger}

\subfloat{
\includegraphics[width=\imsize]
{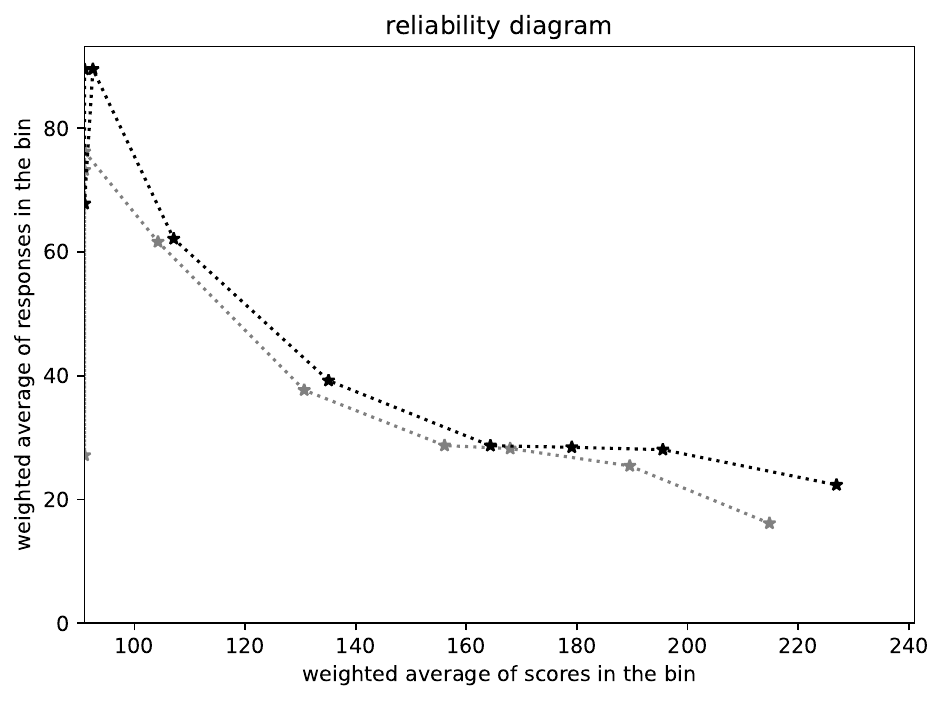}
}
\subfloat{
\includegraphics[width=\imsize]
{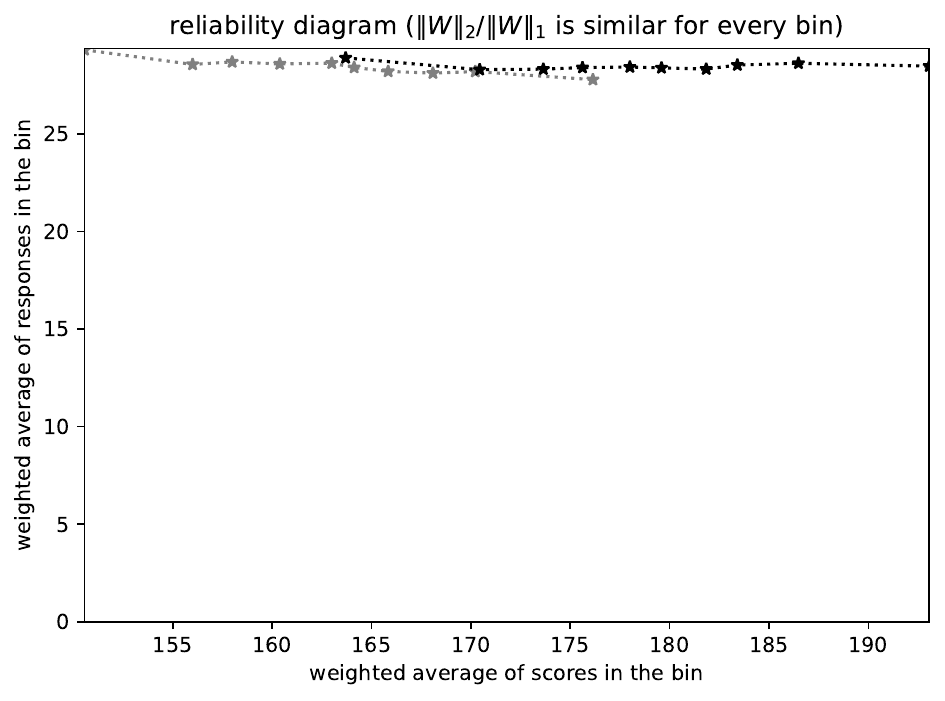}
}

\subfloat{
\includegraphics[width=\imsize]
{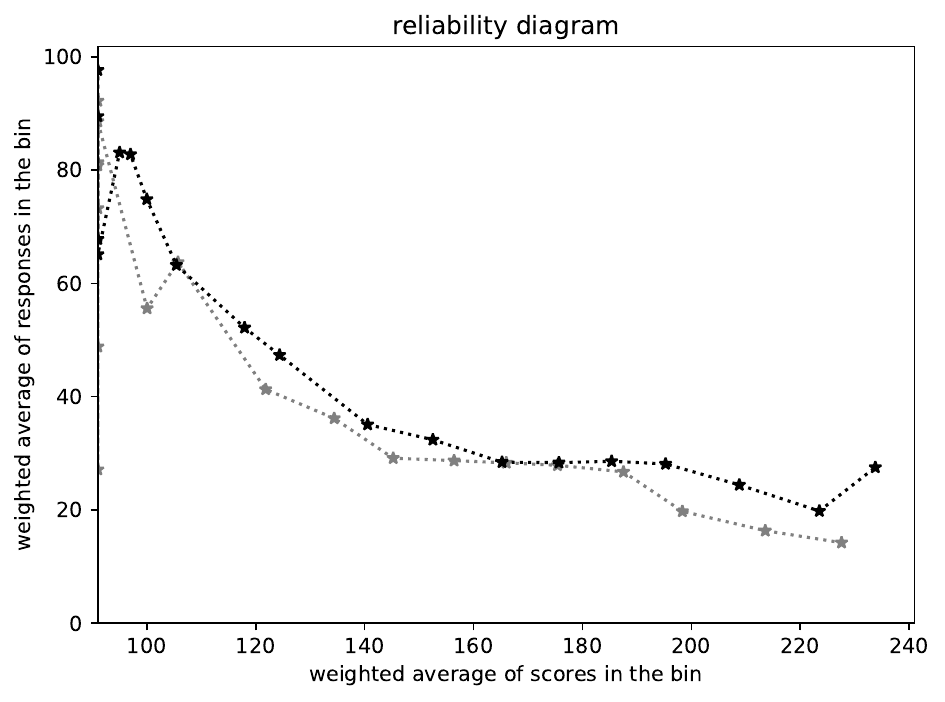}
}
\subfloat{
\includegraphics[width=\imsize]
{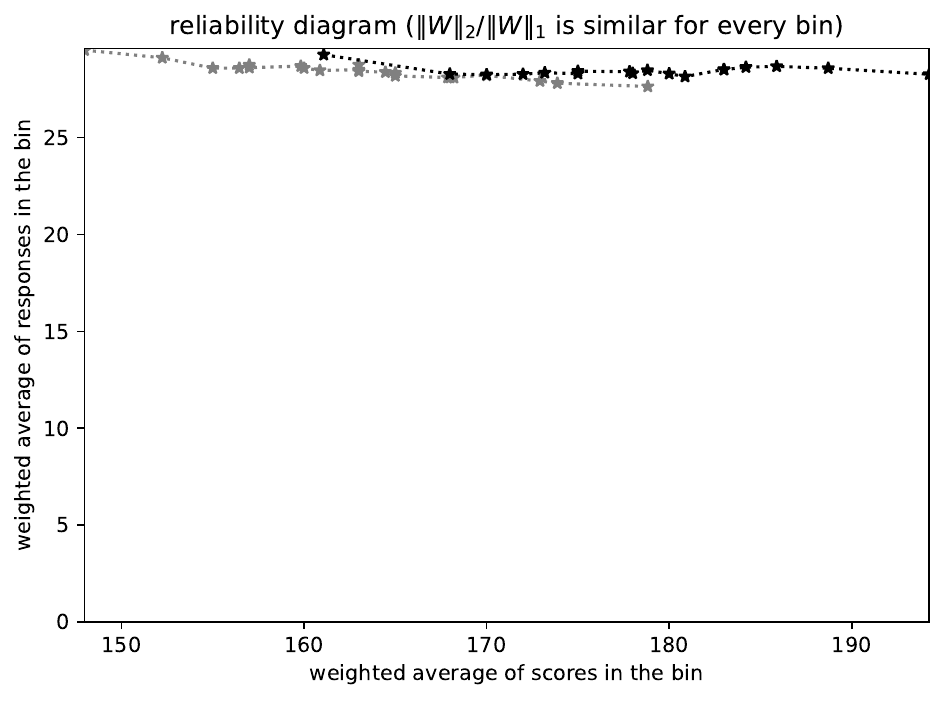}
}

\subfloat{
\includegraphics[width=\imsize]
{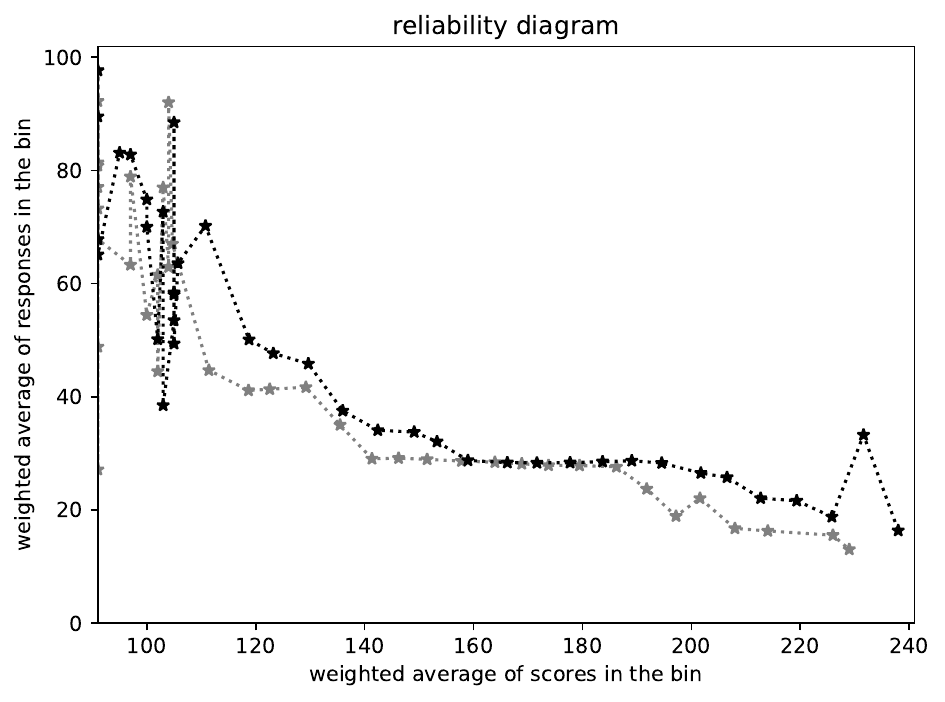}
}
\subfloat{
\includegraphics[width=\imsize]
{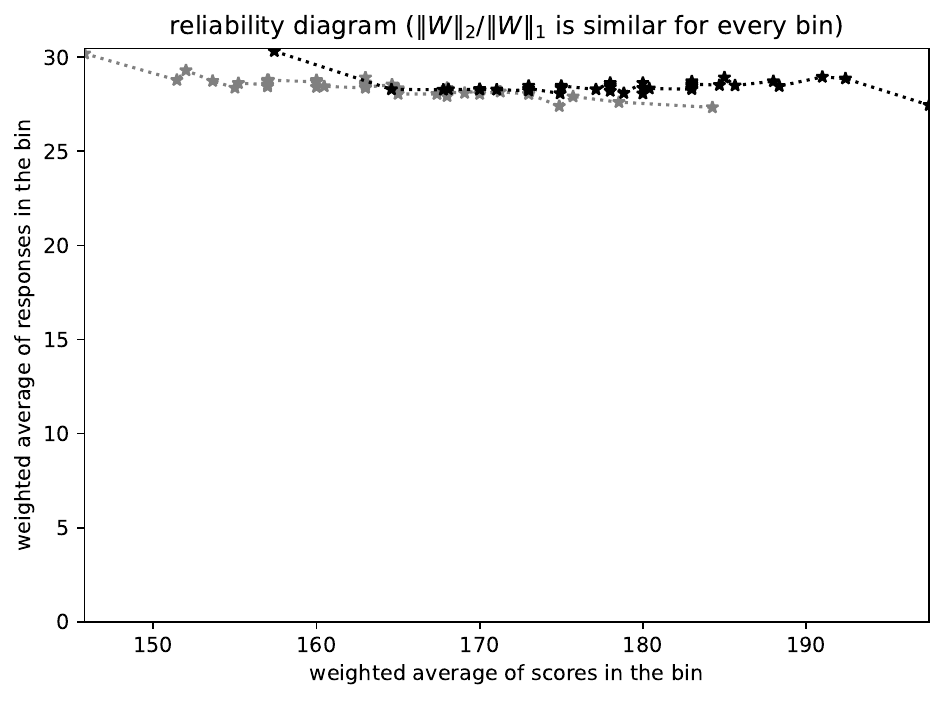}
}
\end{center}
\caption{BMI versus height in centimeters for men compared with women,
with scores perturbed at random using a different random seed (namely, 6789)
than the seeds used in Figures~\ref{men_women1} and~\ref{men_women2}.
The scores for this figure and the other two are heights,
instead of the BMIs used as scores for all other figures;
BMIs are still used here, but as responses rather than scores.
As in Figures~\ref{men_women1} and~\ref{men_women2},
the cumulative plot reveals at a glance much higher BMI for men
than for women who report being very, very short.
The slope of the cumulative graph clearly quantifies how much higher,
which is difficult to assess via the reliability diagrams.}
\label{men_women3}
\end{figure}

\section{Conclusion}
\label{conclusion}

The cumulative statistics originally introduced in order to assess
calibration, bias, and fairness in machine-learned systems extend
to the analysis of biomedical data such as that
of the Behavioral Risk Factor Surveillance System from~\cite{brfss}.
The many advantages of the cumulative approach generalize to biostatistics.

If the covariate is the only confounder, then the cumulative statistics
avoid the paradox of~\cite{simpson} altogether,
by never aggregating the covariate with any binnings
(binning is what causes Simpson's Paradox).
Indeed, graphs of the cumulative differences between subpopulations
display all possible aggregations simultaneously,
with the simple interpretation that the slope of a secant line connecting
two points on the graph is the (weighted) average difference
between the subpopulations over the wide range of values of the covariate
between the two points.

Similarly, the Kuiper metric summarizes the graph of cumulative differences
into a single summary statistic that yields significance tests
with P-values (also known as ``attained significance levels'')
that are valid and converge asymptotically.
In contrast, \cite{arrieta-ibarra-gujral-tannen-tygert-xu}
proved rigorously that the conventional ECE (empirical, estimated, or expected
calibration error) and ICI (integrated calibration index)
are asymptotically inconsistent, diverging as the number of observations
per bin stays bounded while the number of observations becomes large.
While the ECE and ICI pertain directly to assessing calibration
of predicted probabilities and not to the more general comparison
of subpopulations, the present paper illustrates that the advantages
of the Kuiper metric persist in the general setting.

As shown in the results of Section~\ref{results} above,
the cumulative graph is able to display big differences
that are localized to narrow ranges of the covariate
simultaneously with displaying small, noisy, yet significant differences
that require averaging over wide ranges of the covariate to detect.
The conventional reliability diagrams are unable to display
both kinds of differences simultaneously.
This is the basic reason why the classical scalar summary statistics
--- the ECE and the ICI --- diverge as the number of observations becomes large
while the number of observations per bin stays bounded:
the traditional reliability diagrams intrinsically must trade-off
high resolution along the covariate axis for averaging away random noise,
or else trade-off averaging away random noise for high resolution
along the covariate. In contrast, the cumulative graphs
and scalar summary statistics are able to resolve
all statistically significant phenomena without compromise,
while simultaneously admitting simple, intuitive interpretations.

\section*{Acknowledgements}

We would like to thank Kamalika Chaudhuri, Yuval Kluger, and Roy Lederman.

\pagebreak

\appendix
\section{Review of reliability diagrams}
\label{reliability_diagrams}

This appendix summarizes the construction from the traditional method
for assessing calibration of probabilistic predictions and (more generally)
for comparing subpopulations. Assessing calibration of predicted probabilities
is the special case in which the responses for one of the subpopulations are
simply the corresponding scores, where the scores are
the predicted probabilities. The traditional method is known as
the ``reliability diagram'' or occasionally as the ``calibration plot.''
Detailed discussion of reliability diagrams is available
from~\cite{brocker-smith}, \cite{brocker}, or Chapter~9 of~\cite{wilks}.

This appendix follows the notational conventions set at the beginning
of Section~\ref{methods} earlier.

The first step in constructing a reliability diagram is to choose boundaries
for bins used in aggregating the scores and responses.
We denote the edges of the intervals for the bins by the real numbers
$B_1 < B_2 < \dots < B_{p-1}$ together with $B_0 = -\infty$ and $B_p = \infty$,
where $p$ is the number of bins, that is, $p$ is the number of intervals
in a partition of the real line into disjoint intervals
$(B_0, B_1], (B_1, B_2], (B_2, B_3], \dots, (B_{p-2}, B_{p-1}], (B_{p-1}, B_p)$
(again with $B_0 = -\infty$ and $B_p = \infty$).
The other subpopulation can use the same bins or (more usually) different ones
constructed specifically for the other subpopulation.
See below, in the penultimate paragraph of this appendix,
regarding common choices for the bins.

Given the boundaries of the bins, we form the weighted average scores
\begin{equation}
\bar{S}_q = \frac{\sum_{j\,:\,B_{q-1} < S_j \le B_q}
                  S_j \sum_{k=1}^{m_j} W_j^k}
                 {\sum_{j\,:\,B_{q-1} < S_j \le B_q} \sum_{k=1}^{m_j} W_j^k}
\end{equation}
for $q = 1$, $2$, \dots, $p$, in the case of paired samples
or when considering the full population.
When the subpopulation is instead specified by ordered pairs of indices
$i_1^0$, $i_2^0$, \dots, $i_{n_0}^0$, we form the weighted average scores
\begin{equation}
\bar{S}_q = \frac{\sum_{j\,:\,B_{q-1} < S_{\left(i_j^0\right)_1} \le B_q} \,
                  S_{\left(i_j^0\right)_1} \,
                  W_{\left(i_j^0\right)_1}^{\left(i_j^0\right)_2}}
                 {\sum_{j\,:\,B_{q-1} < S_{\left(i_j^0\right)_1} \le B_q}
                  W_{\left(i_j^0\right)_1}^{\left(i_j^0\right)_2}}
\end{equation}
for $q = 1$, $2$, \dots, $p$.

We also form the weighted average responses
\begin{equation}
\bar{R}_q = \frac{\sum_{j\,:\,B_{q-1} < S_j \le B_q}
                  \sum_{k=1}^{m_j} R_j^k \, W_j^k}
                 {\sum_{j\,:\,B_{q-1} < S_j \le B_q} \sum_{k=1}^{m_j} W_j^k}
\end{equation}
for $q = 1$, $2$, \dots, $p$, in the case of paired samples
or when considering the full population.
And, when the subpopulation is instead specified by ordered pairs of indices
$i_1^0$, $i_2^0$, \dots, $i_{n_0}^0$, we form the weighted average responses
\begin{equation}
\bar{R}_q = \frac{\sum_{j\,:\,B_{q-1} < S_{\left(i_j^0\right)_1} \le B_q} \,
                  R_{\left(i_j^0\right)_1}^{\left(i_j^0\right)_2} \,
                  W_{\left(i_j^0\right)_1}^{\left(i_j^0\right)_2}}
                 {\sum_{j\,:\,B_{q-1} < S_{\left(i_j^0\right)_1} \le B_q}
                  W_{\left(i_j^0\right)_1}^{\left(i_j^0\right)_2}}
\end{equation}
for $q = 1$, $2$, \dots, $p$.

The reliability diagram consists of plotting the ordered pairs
$(\bar{S}_1, \bar{R}_1)$, $(\bar{S}_2, \bar{R}_2)$, \dots,
$(\bar{S}_p, \bar{R}_p)$ in black, together with plotting in gray
the corresponding ordered pairs associated with the other subpopulation
(or with the full population, when comparing a subpopulation
to the full population).

There are two common choices for bins. The classical choice is to require
the widths of the finite-width bins to be identical, that is,
$B_2 - B_1 = B_3 - B_2 = \dots = B_{p-1} - B_{p-2}$.
However, most literature about reliability diagrams points to the advantages
of choosing the bins such that the error bars for the bins are roughly equal.
Equalizing the error bars is possible by selecting the widths of the bins
such that the ratio of the sum of the squares of the weights for observations
in a given bin to the square of the sum of the weights in that bin is roughly
the same for every bin. Remark~5 of~\cite{tygert_full} provides
a simple algorithm for calculating such widths for the bins.
Of course, when all weights are the same, then this results in each bin
containing a similar number of observations,
thus averaging away noise about as much as in every other bin.

Figures~\ref{angina_heart-attack}--\ref{hiv_kidney3}
display reliability diagrams in which the ratio of the sum of the squares
of the weights for observations in a given bin to the square of the sum
of the weights in that bin is similar for every bin.
The title of each of these reliability diagrams is commensurate,
``reliability diagram ($\|W\|_2 / \|W\|_1$ is similar for every bin).''
Figures~\ref{men_women1}--\ref{men_women3}
display both reliability diagrams in which the widths of the finite-width bins
are all the same and reliability diagrams in which the ratio of the sum
of the squares of the weights for observations in a given bin to the square
of the sum of the weights in that bin is similar for every bin.
The titles for the former are the shorter, ``reliability diagram,''
while the titles for the latter are longer,
``reliability diagram ($\|W\|_2 / \|W\|_1$ is similar for every bin).''

\bibliography{paper}
\bibliographystyle{authordate1}

\end{document}